\documentclass[showpacs,superscriptaddress,10pt,nofootinbib]{revtex4}
\usepackage[utf8x]{inputenc}
\usepackage{slashed}
\usepackage{amssymb,amsmath}
\usepackage[dvips,final]{graphicx}
\usepackage{url}
\usepackage{subfigure}
\usepackage{textcomp}
\usepackage{bm}
\usepackage{dcolumn}
\usepackage[usenames]{color}
\usepackage{amsthm}
\usepackage{amsopn}
\usepackage{ulem}
\usepackage{cancel}
\usepackage{float}
\usepackage{epsfig}
\usepackage{braket}
\usepackage[title]{appendix}
\usepackage{amsopn}
\usepackage{hyperref}
\usepackage{soul}

\def\comment#1{}

\def\beq{\begin{equation}}
\def\eeq{\end{equation}}
\def\bea{\begin{eqnarray}}
\def\eea{\end{eqnarray}}

\newcommand{\quantities}[1]{%
	\begin{tabular}{@{}c@{}}\strut#1\strut\end{tabular}%
}

\input epsf.tex
\def\comment#1{}


\begin{document}

\title{CMB circular and B-mode polarization from new interactions}

\author{Nicola Bartolo}
\email[]{nicola.bartolo@pd.infn.it}

\affiliation{Dipartimento di Fisica e Astronomia \textquotedblleft G. Galilei\textquotedblright, Universit\`{a} degli Studi di Padova, via Marzolo
8, I-35131, Padova, Italy}
\affiliation{INFN Sezione di Padova, via Marzolo 8, I-35131, Padova, Italy}
\affiliation{INAF-Osservatorio Astronomico di Padova, Vicolo dell'Osservatorio 5, I-35122 Padova, Italy}

\author{Ahmad Hoseinpour}
\email[]{ahmad.hoseinpour@ph.iut.ac.ir}

\affiliation{Department of Physics, Isfahan University of Technology, Isfahan
84156-83111, Iran}
\affiliation{ICRANet-Isfahan, Isfahan University of Technology, 84156-83111, Iran}

\author{Sabino Matarrese}
\email[]{sabino.matarrese@pd.infn.it }

\affiliation{Dipartimento di Fisica e Astronomia \textquotedblleft G. Galilei\textquotedblright, Universit\`{a} degli Studi di Padova, via Marzolo
8, I-35131, Padova, Italy}
\affiliation{INFN Sezione di Padova, via Marzolo 8, I-35131, Padova, Italy}
\affiliation{INAF-Osservatorio Astronomico di Padova, Vicolo dell'Osservatorio 5, I-35122 Padova, Italy}
\affiliation{Gran Sasso Science Institute, viale F. Crispi 7, I-67100, L'Aquila, Italy}

\author{ Giorgio Orlando}
\email[]{giorgio.orlando@phd.unipd.it}

\affiliation{Dipartimento di Fisica e Astronomia \textquotedblleft G. Galilei\textquotedblright, Universit\`{a} degli Studi di Padova, via Marzolo
8, I-35131, Padova, Italy}
\affiliation{INFN Sezione di Padova, via Marzolo 8, I-35131, Padova, Italy}

\author{Moslem Zarei}
\email[]{m.zarei@cc.iut.ac.ir}

\affiliation{Department of Physics, Isfahan University of Technology, Isfahan
84156-83111, Iran}
\affiliation{ICRANet-Isfahan, Isfahan University of Technology, 84156-83111, Iran}

\date{\today}

\date{\today}

 \begin{abstract}
Standard models describing the radiation transfer of the cosmic microwave background (CMB) through Compton scattering predict that cosmological scalar perturbations at linear order are not able to source V and B polarization modes. In this work we investigate the possibility that such CMB polarization modes are generated even in the presence of linear scalar perturbations only. We provide a general parametrization of the photon-fermion forward-scattering amplitude and compute mixing terms between different CMB polarization modes. We discuss different general extensions of Standard Model interactions which violate discrete symmetries, while preserving the combination of charge conjugation, parity and time reversal. We show that it is possible to source CMB circular polarization by violating parity and charge conjugation symmetries. Instead, B-mode generation is associated to the violation of symmetry for time-reversal. Our results provide a useful tool to constrain new physics using CMB data.
 \end{abstract}

%\pacs{13.15.+g, 98.80.Es, 98.70.Vc.}

\maketitle
%\newpage

\section{Introduction}

CMB radiation represents a crucial observational tool of modern cosmology. The standard models describing the radiation transfer of the CMB from the recombination epoch until today predict the presence of some level of linear polarization, the so-called E- and B-modes, which have been widely studied and reviewed in the literature (see e.g. Refs. \cite{Kosowsky:1994cy, Seljak:1996is,Zaldarriaga:1996xe, Kamionkowski:1996ks, Kamionkowski:1996zd, Hu:1997, Dodelson:2003, dodelson:2017}). This is the result of the Compton scattering between CMB photons and electrons and gravitational redshift, induced by cosmological perturbations of the metric. Instead, the generation of CMB circular polarization (the so-called V-mode) is usually not considered, because the electron-photon Compton scattering cannot generate it at the classical level.

However, some models have been proposed that can lead to the generation of CMB circular polarization. One possible way is via Faraday conversion of the linear polarization generated at the surface of last scattering by various sources of cosmic birefringence (see e.g. Refs. \cite{Montero:2018,Marc:2018} for a recent review).
For instance, in Refs. \cite{Kosowsky:1996yc,Giovannini:2002,Cooray:2003,Scoccola:2004ke,Campanelli:2004pm,Giovannini:2008,Zarei:2010,De:2015, Ejlli:2016, Ejlli:2017}, V-mode formation due to magnetic fields is discussed. In Refs. \cite{Motie:2012,Mohammadi:2014,Sawyer:2015,Sadegh:2018}, V-mode formation due to photon-photon interactions via Heisenberg-Euler interaction is considered. V-mode generation due to interactions coming from extensions of QED is studied, in particular, in Refs. \cite{Colladay:1998,Alexander:2009,Zarei:2010}, where Lorentz-violating operators are considered.  In  Ref. \cite{Finelli:2009}, it is shown that a cosmological pseudoscalar field may generate circular polarization in the CMB, while in Ref. \cite{Alexander:2017} it is shown that V-mode generation can be obtained in axion inflation. Moreover, in Refs. \cite{Mohammadi:2013, Xue:2014}, it is shown that forward scattering between CMB photons and neutrinos can source V-modes through Standard Model interactions. Also, in Ref.~\cite{Bartolo:2018}, forward scattering between photons and gravitons is shown to lead to circular polarization, under some conditions. In Ref. \cite{Vahedi:2018}, circular polarization of CMB photons via their Compton scattering with polarized cosmic electrons is considered. In Ref. \cite{Inomata:2018}, it is shown that V-modes in the CMB may arise from primordial vector and tensor perturbations. In particular, in Refs. \cite{Kamionkowski:2018, Alexander:2018}, the case of chiral gravitational waves is considered.

Despite the fact that CMB circular polarization has not been explored so much up to now, these examples show how its detection might reveal interesting phenomena occurring in the evolution of the universe.

Most of the mechanisms to produce V-modes proposed in recent years are based on the forward scattering of CMB photons by a target. In fact, the generation of V-modes depends on the refractive index of the material, which is related to the forward scattering amplitude $M_{for}$ of a fundamental process (see e.g. Ref. \cite{Karl:1975}). In particular, circular polarization is generated when the refractive index of the left-handed (LH) waves differs from the refractive index of the right-handed (RH) ones. As a result, the existence of a nonvanishing V-mode implies that $M^R_{for}\neq M^{L}_{for}$.

In the language of quantum mechanics, the forward scattering amplitude of a beam of radiation $\gamma$ and a target $A$ is given as $M^{R,L}_{for}=\langle\gamma,A|\hat{\mathcal{O}}|\gamma,A\rangle_{R,L}$, where $|\gamma,A\rangle$ represents the quantum state of the target and of the beam, and $\mathcal{O}$ is the interaction operator. The condition $M^R_{for}\neq M^L_{for}$ is satisfied either if (i) the state of target $| A\rangle$ is not a parity eigenstate, namely $P| A\rangle\neq \pm |A\rangle$ or if (ii) $\mathcal{O}$ is not invariant under parity transformation, namely $P\mathcal{O}P^{-1}\neq\mathcal{O}$. There are several ways in which the first condition can be met. For example, forward scattering of photons with a background of particles can produce V-modes when the power-spectrum of this background violates parity symmetry. Instead, Compton scattering in the presence of a magnetic field is an example of the second condition. Historically, Ref.~\cite{Lue:1998mq} was the first literature that pointed out the possibility to use the CMB to search for parity violating interactions.

From the observational point of view, CMB circular polarization is not excluded. As an example, the SPIDER collaboration has recently provided new constraints on the Stokes parameter V at 95 and 150 GHz, by observing angular scales corresponding to  $33< \ell <307$ \cite{SPIDER}. The constraints on the circular polarization power-spectrum $\ell(\ell + 1) C^{\ell}_{VV} /(2\pi)$ are reported in a range from 141 $\mu {\rm K}^2$ to 255 $\mu {\rm K}^2$ at 150 GHz for a thermal CMB spectrum. Also, in Ref.~\cite{King:2016}, some interesting detection prospects are discussed.

In this work, we will study V-mode polarization generation in the CMB radiation from its direct coupling with linear polarization states induced by the forward scattering of photons with generic fermions at or after the recombination epoch. In particular, we will assume a completely general photon-fermion interaction which may also go beyond QED, but still preserving the combination of charge conjugation, parity and time reversal (CPT), which up to now is observed to be an exact symmetry of nature at a fundamental level. In order to do so, we will use a generic parametrization of the photon-fermion scattering amplitude which follows only by the imposition of gauge-invariance (see e.g. Refs. \cite{Prange,Drechsel:2002ar,Lifshitz}). Moreover, we will work in the so-called ``quantum Boltzmann equation" formalism (see e.g. Refs. \cite{Kosowsky:1994cy, Alexander:2009,Zarei:2010,Mohammadi:2013,Mohammadi:2014,Mohammadi:2015,Batebi:2016ocb,Sadegh:2018,Bartolo:2018, Shakeri:2018}) for computing the time evolution of CMB polarization. It is possible to show that this formalism is equivalent at lowest order in scattering kinematics to the classical radiation transfer; hence it provides a more general framework to work with. Our results fall either in category (i) or (ii) (or both) as defined above, according to the different cases considered.

We will show that V-modes can be produced by forward scattering for a generic interaction preserving all the C (charge conjugation), P (parity), and T (time-reversal) discrete symmetries, if the stress tensor of the fermion contains anisotropies. In addition, we will show that V-modes can be sourced also from an interaction violating C and P symmetries, but preserving the CP combination. In this case, together with the anisotropies in the fermionic stress tensor, we need the fermion to interact with the photon only in the L- or R-handed helicity state, like the L-handed neutrino in the Standard Model interactions. In particular, this last case confirms and generalizes the results found in Ref.~\cite{Mohammadi:2013}. We will also analyze the cases in which C, T and P, T symmetries are violated individually, while preserving, respectively, the combinations CT and PT. We will show that in these cases it is impossible to generate V-modes by forward scattering, but we can have formation of CMB B-modes. In particular, in the case of a generic interaction which violates P, T symmetries, it is possible to generate B-modes with no conditions on the fermions the photons interact with, while in the case in which C, T are violated we need the fermion to be in the  L- or R-handed helicity state. All these conclusions, which represent the main results of this paper, are summarized in Table \ref{table}.

Thus, our general study shows that the forward scattering term produced by interactions beyond the Standard Model may produce V-modes in the CMB and, at the same time, can provide an additional source of B-modes. Our final Boltzmann equations are expressed in terms of unknown free parameters. Thus, in the future we could use our general approach to put constraints on new physics using CMB data.

The paper is organized as follows. In Sec. \ref{2}, we will introduce a set of equations and useful notations and results that we will use in this work.
In Sec. \ref{3}, we will see a generic way to parametrize the photon-fermion scattering amplitude and apply this parametrization in specific cases. In Sec. \ref{4},  we will derive the general form of the forward scattering term in the photon-fermion interaction. In Sec. \ref{5}, we will study general interactions generating V-mode polarization in the CMB. In Sec. \ref{Bmode}, we will see some cases where also CMB B-modes can be sourced. In Sec. \ref{7}, we will comment about the difference in the results, by considering the fermion as a Majorana particle, instead of a Dirac particle. Finally, in Sec. \ref{8}, we will present our main conclusions.

%%%%%%%%%%%%%%%%%%%%%%%%%%%%%%%%%%%%%%%%%%%%%%%%%%%%%%%%%%%%%%%%%%%%%%%%%%%%%%%%%%%%%%%%%%%%%%%%%%%%%%%%%%%%%%%%%%%%%%%%

\section{The time evolution of Stokes parameters}\label{2}

 The intensity and polarization of CMB anisotropies are completely characterized by a $2\times 2$ polarization matrix\footnote{When we refer to the Stokes parameters, we take only the fluctuations over the respective mean value.}
\bea
 \rho_{i j}=\frac{1}{2}\left(\begin{array}{cc}
             I+Q& U-iV \\
             U+iV & I-Q \\
                 \end{array}
        \right)\,,
        \label{density matrix}
\eea
where $I$, $Q$, $U$, and $V$ are the so-called
Stokes parameters, satisfying the inequality $I^2\geq Q^2+U^2+V^2$ \cite{Kosowsky:1994cy}. The components of the polarization matrix $\rho_{ij}$ satisfy the relations
 $\rho_{ii}=1$ and $\rho_{ij}=\rho^{\ast}_{ij}$ or, better to say, the $\rho_{ij}$ matrix is Hermitian. Consequently, the diagonal components $\rho_{11}$ and $\rho_{22}$ are real (with $\rho_{11}+\rho_{22}=1$), while $\rho_{21}=\rho^{\ast}_{12}$.

For unpolarized CMB radiation $Q = U = V = 0$, and the parameter $I$ describes the overall radiation intensity. The $Q$ and $U$ Stokes parameters represent the linear polarization of the CMB. In particular, taking two orthogonal (x,~y) axes on the polarization plane, the Q-mode gives the difference in intensity between CMB photons with polarization vectors along the x and y axes, respectively, while the U-mode gives the difference in intensity between CMB photons with a polarization vector along axes rotated by 45 degrees with respect to the x and y axes. Finally, the $V$-mode describes the CMB circular polarization or, better to say, it gives the difference in intensity between the two circular polarization modes of CMB radiation.

The generation and evolution of CMB intensity and polarization can be characterized through the quantum Boltzmann equation \cite{Kosowsky:1994cy}
\bea
(2\pi)^3 \delta^{(3)}(0)(2k^0)
\frac{d\rho_{ij}(\mathbf{k})}{dt} = i\left\langle \left[H_I
(t),\mathcal{D}_{ij}(\mathbf{k})\right]\right\rangle-\frac{1}{2}\int _{-\infty}^{\infty} dt\left\langle
\left[H_I(t),\left[H_I
(0),\mathcal{D}_{ij}(\mathbf{k})\right]\right]\right\rangle \, , \label{Boltzmann equation}
\eea
where $\langle \cdot\cdot\cdot \rangle$ denotes the expectation value of operators, $\mathcal{D}_{ij}(\mathbf{k})=a^\dagger_{i}(\mathbf{k})a_{j}(\mathbf{k})$ is the photon number operator, $a^{\dag}$ and $a$ are the photon creation and annihilation operators, respectively. The effective interaction Hamiltonian $H_I$ is defined through the expansion of the S matrix up to second order as
\beq
S^{(2)}=-i\int_{-\infty}^{\infty}\,dt\, H^{(2)}(t) \label{intH2} \; ,
\eeq
where $H_I$ is the component of $H^{(2)}$ that describes the Compton scattering between CMB photons and other particles. The first term on the right-hand side of Eq. \eqref{Boltzmann equation} is the so-called {\it forward scattering} term, while the second term is the so-called {\it damping} or {\it nonforward scattering} term. In this work, we will focus on the forward scattering term which is able to generate couplings between different polarization states.\footnote{This is the same physical mechanism that generates neutrino flavor mixings; see e.g. Ref. \cite{Sigl:1992fn}.} In fact, Eq. \eqref{Boltzmann equation} is derived adopting a perturbative approach so that increasing powers of the interaction Hamiltonian $H_I(t)$ reduce the strength of the corresponding term. For this reason, in any fundamental interaction in perturbative regime in which the forward scattering term is nonzero, a priori it is expected to give the relevant physical effects on the CMB polarizations. Of course, this is not the case of the standard QED interaction between photons and electrons where such a forward scattering term vanishes (see e.g. \cite{Kosowsky:1994cy}), and all the relevant effects arise from the damping term only. This is also of the main reasons why in this paper we focus on the forward scattering term only.

The Stokes parameters inside the polarization matrix can be expanded in terms of a spin-weighted basis as
\beq
I(\mathbf{\hat{k}})=\sum_{\ell, m}a^{I}_{\ell m}Y_{\ell m}(\mathbf{\hat{k}})~,
\eeq
\beq
V(\mathbf{\hat{k}})=\sum_{\ell, m}a^{V}_{\ell m}Y_{\ell m}(\mathbf{\hat{k}})~,
\eeq
\beq
P^{\pm}(\mathbf{\hat{k}})=(Q\pm iU)(\mathbf{\hat{k}})=\sum_{\ell, m}a_{\pm2,\ell m}\,_{\pm 2}\!Y_{\ell m}(\mathbf{\hat{k}})\, ,
\eeq
where $\mathbf{\hat{k}}$ denotes the photon direction. Moreover, using the spin raising and lowering operators $\eth$ and $\bar{\eth}$, we get
\beq \label{E}
{E}(\mathbf{\hat{k}})=-\frac{1}{2}\left[\bar{\eth}^2 P^{+}(\mathbf{\hat{k}})+\eth^2 P^{-}(\mathbf{\hat{k}})\right]~,
\eeq
\beq \label{B}
{B}(\mathbf{\hat{k}})=\frac{i}{2}\left[\bar{\eth}^2 P^{+}(\mathbf{\hat{k}})-\eth^2 P^{-}(\mathbf{\hat{k}}) \right]~,
\eeq
where we have introduced the so-called E and B polarization modes. These modes offer an alternative description of CMB linear polarization which, differently from Q-and U-modes, is invariant under a rotation of the polarization plane. In the following, we will use a description of the radiation transfer both in terms of Q- and U-modes and E- and B-modes.

The standard Boltzmann equations in the presence of only linear scalar perturbations are given by \cite{Kosowsky:1994cy}
\beq
\frac{d}{d \eta}I^{(S)} +iK\mu \, I^{(S)}+4[\psi'-i K \mu \phi]
= \tau'\Big[I^{(S)} -
{I}^{0(S)} +4\mu v_B -{1\over 2}P_2(\mu)\,\Pi\Big] \, , \label{Boltzmann}
 \eeq
\beq
\frac{d}{ d \eta}{P}^{\pm (S)} +i K\mu {P}^{\pm (S)} = \tau'\Big[
{P}^{\pm (S)} +{1\over 2} [1-P_2(\mu)]\, \Pi\Big] \, ,\label{EqP}
\eeq
\beq
\frac{d}{d \eta}{V}^{ (S)} +i K\mu{V}^{ (S)} = \tau'\left[
{V}^{(S)} -{3\over 2} \mu {V}^{1(S)} \right] \, ,
\label{eqV1}
\eeq
 where $\mathbf{K}$ denotes the Fourier conjugate of $\mathbf{x}$ and we are in a coordinate system where $\mathbf{K} \parallel \hat{\mathbf{z}}$ axis. Here $\eta$ is the conformal time, the prime denotes differentiation with respect to conformal time, $\psi$ and $\phi$ are scalar cosmological (gravitational potential) perturbations, $v_B$ is the electrons average velocity,
$\mu=\hat{k} \cdot \hat{K}  =\cos\theta$,  $P_\ell(\mu)$ is the Legendre polynomial of rank $\ell$ and $\Pi\equiv {I}^{2(S)}+{P}^{2(S)}-{P}^{0(S)}$, $P$ being the strength of the polarization field. The quantities $I^{\ell(S)}$, $P^{\ell(S)}$ and $V^{\ell(S)}$ represent the $\ell$-th order terms in the Legendre polynomial expansion of the corresponding modes\footnote{We adopt the convention
	\beq
I^{\ell(S)} = \int_{-1}^1 \frac{d\mu'}{2} I(k, \mu') P_\ell(\mu') \,,
	\eeq
and an analogous expression for $V^{\ell(S)}$ and $P^{\ell(S)}$.}. Finally, one defines the optical depth $\tau (\eta)$ as
\begin{equation}\label{optical}
\tau'(\eta)=- a(\eta)n_Bx_e\sigma_T\,,\,\,\,\,\,\,\,\,\quad \,\,\,\,\,\,\,\tau(\eta)= - \int_\eta^{\eta_0} \tau'(\eta')\, d\eta' \, ,
\end{equation}
where $n_B$ is the electron density, $x_e$ is the ionization
fraction and $\sigma_T$ is the Thomson cross section. For more details about the derivation of Eqs. \eqref{Boltzmann}, \eqref{EqP} and \eqref{eqV1}, see Refs. \cite{Kosowsky:1994cy,Hu:1997, Dodelson:2003}.\footnote{We note that the Boltzmann equations do not coincide between the different references. This is due to the fact that each reference uses its own formal conventions. In our results, we have followed the conventions of Ref. \cite{Kosowsky:1994cy}.}

In particular, it is possible to show that Eq. \eqref{EqP} admits the general integral solution
\begin{eqnarray}
{P}^{\pm (S)}
(\eta_0,K,\mu)&=&{3 \over 4}(1-\mu^2)\int_0^{\eta_0} d\eta\,
e^{iK(\eta - \eta_0)\mu  -\tau}\,\tau'\,\Pi(\eta, K)\, .\label{Boltzmann3}
\end{eqnarray}
Since scalar perturbations are invariant under rotations and so axially symmetric around $\hat{\mathbf z}$, we get $P^{+}= P^{-}$; thus, $U^{(S)}= 0$ and scalar perturbations source only Q-modes. Moreover, in this case ({\it i.e.} for scalar perturbations) the spin raising and lowering operators act like (see e.g. Ref. \cite{shiraishi:2013})
\beq \label{P_to_EB}
\bar{\eth}^{2}\, {P}^{\pm (S)} = \eth^2 P^{\pm(S)}=  \partial_{\mu}^{2}[(1-\mu^{2})\,\, {P}^{\pm (S)}(\eta_0, K,\mu)]~.
\eeq
Therefore, using the definitions \eqref{E} and \eqref{B} we get
\beq
{E}^{ (S)}
(\eta_0, K,\mu)=-{3 \over 4 }\int_0^{\eta_0} d\eta\,e^{-\tau}\tau'\Pi(\eta, K)
\partial_{\mu}^2\left[(1-\mu^2)^2 e^{iK(\eta - \eta_0) \mu }\right]~,
\eeq
and
\beq
{B}^{ (S)}
(\eta_0, K,\mu)=0 \, .
\eeq
This is the well-known result that linear scalar perturbations cannot source B-mode polarization. In fact, it is well known that the B-mode polarization in the CMB is generated mainly by weak gravitational lensing and by tensor perturbations  \cite{Dodelson:2003, dodelson:2017}. Alternatively, a small amount of B-modes can be generated also by second-order vector and tensor modes sourced by scalar perturbations (see e.g. Refs. \cite{Mollerach:2004, Fidler:2014oda}).

Moreover, since in the right-hand side of Eq. \eqref{eqV1} the Stokes parameter $V$ has no source terms in the case of vanishing initial conditions for V, neither  V-modes can arise with linear scalar perturbations only.

From the next section we will start to study the general conditions for generating both V- and B-modes in the presence of only linear scalar perturbations through their direct coupling with E-modes generated by the photon-fermion forward scattering. We will first write down a very general form for the photon-fermion scattering  amplitude (Sec.~\ref{3}), and then we will apply it to the forward scattering contribution in Eq.~(\ref{Boltzmann equation}) (in Sec.~\ref{4}).

%%%%%%%%%%%%%%%%%%%%%%%%%%%%%%%%%%%%%%%%%%%%%%%%%%%%%%%%%%%%%%%%%%%%%%%%%%%%%%%%%%%%%%%%%%%%%%%%%%%%%%%%%%%%%%%%%%%%%%%%
%%%%%%%%%%%%%%%%%%%%%%%%%%%%%%%%%%%%%%%%%%%%%%%%%%%%%%%%%%%%%%%%%%%%%%%%%%%%%%%%%%%%%%%%%%%%%%%%%%%%%%%%%%%%%%%%%%%%%%%%

\section{General form of the photon-fermion scattering amplitude} \label{3}

\begin{figure}[htp]
  \includegraphics[width=1.5in]{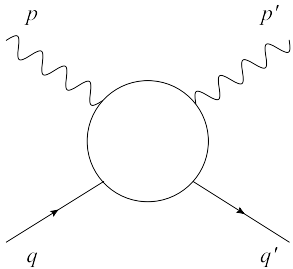}\\
  \caption{Figurative representation of photon-fermion interaction.
  }\label{fig1}
\end{figure}

We are interested in the Compton scattering of a photon by a fermion (Fig. \ref{fig1})
\beq
\gamma(p)+f(q)\rightarrow \gamma(p')+f(q')~,
\eeq
where $p\, (p')$ is the initial (final) momentum of the photon and $q\,(q')$ is the initial (final) momentum of the fermion. It is possible to construct the invariant amplitude of this process using a general method. The amplitude of such a process can be written in the form \cite{Prange,Drechsel:2002ar,Lifshitz}
\beq
M_{fi}=  F^{\lambda\mu}\epsilon^{s'\ast}_{\lambda}\epsilon^s_{\mu}~. \label{amplitude0}
\eeq
where $\epsilon^{s}_{\mu}$ and $\epsilon^{s'}_{\nu}$ are the polarization vectors of incoming and outgoing photons and $s,s'=1,2$ label the physical transverse polarization of the photons.
Gauge-invariance requires $\epsilon^{s}\cdot p=\epsilon^{s'}\cdot p'=0$. Moreover, the rank-2 tensor $F^{\mu\nu}$, which is called ``Compton tensor", must satisfy the conserved current condition $p_{\mu}F^{\mu\nu}=p'_{\nu}F^{\mu\nu}=0$, as a consequence of gauge-invariance. It is possible to provide a general parametrization of $F^{\mu\nu}$ satisfying the previous condition from the linear combination of basis vectors defined below.

We first construct a general form for the Compton tensor $F^{\mu\nu}$ and then study its parity conserving and parity violating aspects. Using the procedure of Refs. \cite{Prange,Drechsel:2002ar,Lifshitz}, we can write
\bea
F^{\mu\nu}&=&G_0\left(\hat{e}^{(1)\mu}\hat{e}^{(1)\nu}+\hat{e}^{(2)\mu}\hat{e}^{(2) \nu}\right)
+G_1\left(\hat{e}^{(1) \mu}\hat{e}^{(2) \nu}+\hat{e}^{(2) \mu}\hat{e}^{(1) \nu}\right)
+G_2\left(\hat{e}^{(1) \mu}\hat{e}^{(2) \nu}-\hat{e}^{(2) \mu}\hat{e}^{(1) \nu}\right)\nonumber \\&&
+G_3\left(\hat{e}^{(1) \mu}\hat{e}^{(1) \nu}-\hat{e}^{(2) \mu}\hat{e}^{(2) \nu}\right)\,,\label{amplitude1}
\eea
where $G_i$ are invariant functions and $e^{(1)}$ and $e^{(2)}$ are two 4-vectors satisfying the orthogonality condition $\hat{e}^{(1)}\cdot \hat{e}^{(2)}=0$.
In order to construct these two vectors, we have to use only the kinematic variables $p$, $p'$, $q$, and $q'$ and define a system of orthogonal vector basis of the form
\beq
Q^{\lambda}=(q^{\lambda}+q'^{\lambda})-\frac{P^{\lambda}}{P^{2}}\:(q+q')\cdot P\,,\label{Q1}
\eeq
\beq
P^{\lambda}=p^{\lambda}+p'^{\lambda}\,,
\eeq
\beq \label{N}
N^{\lambda}=\epsilon^{\lambda\mu\nu\rho}Q_{\mu}t_{\nu}P_{\rho}\,,
\eeq
where $t^{\lambda}$, for the tree-level contribution to the scattering amplitude, is given by
\beq
t^{\lambda}=q^{\lambda}-q'^{\lambda}=p'^{\lambda}-p^{\lambda}\,.
\eeq
A possible choice of the normalized $\hat{e}^{(1)}$ and $\hat{e}^{(2)}$ four-vectors is given by (see e.g. \cite{Prange})
\beq \label{def}
\hat{e}^{(1)\lambda}=\frac{N^{\lambda}}{\sqrt{-N^2}}\,,
\eeq
and
\beq
\hat{e}^{(2)\lambda}=\frac{Q^{\lambda}}{\sqrt{-Q^2}}~.
\eeq
From these definitions it is easy to verify the conserved current condition as
\beq
 (P_{\nu}+t_{\nu})F^{\mu\nu}=(P_{\mu}-t_{\mu})F^{\mu\nu}=0\, .
\eeq
In this paper, we are interested in the forward scattering limit in which $t^{\lambda}=0$ and $P^2=4p^2=0$. Under this condition, $N^{\lambda}$ vanishes and the second term in $Q^{\lambda}$ becomes singular. Therefore, $\hat{e}^{(1)\lambda}$ and $\hat{e}^{(2)\lambda}$ are not well-defined. In order to overcome these problems, we firstly change the normalization in $\hat{e}^{(2)\lambda}$ as
\beq \label{e2}
\hat{e}^{(2)\lambda}=\frac{Q^{\lambda}}{\sqrt{-4q^2}}=\frac{Q^{\lambda}}{\sqrt{-4m^2_f}}~,
\eeq
by noting that the second term in $Q^{\lambda}$ does not contribute to the amplitude. Second, we introduce a new general quantity $\Delta^{\lambda}$ replacing $t^{\lambda}$ in Eq. \eqref{N}. This quantity $\Delta^{\lambda}$ has to be expressed in terms of kinematic variables and invariants of the interaction. However, in the forward scattering limit, any linear combination of the photon and electron four-momenta $p^{\mu}$ and $q^{\mu}$ leads to a negligible value of $\hat{e}^{(1)\lambda}$ when doing the contractions with the Levi-Civita pseudotensor in Eq. \eqref{N}. Hence, $\Delta^{\lambda}$ has to be given only in terms of scalar  invariant quantities. Thus, the only possibility to define $\Delta^{\lambda}$ reads
\beq \label{delta}
\Delta^{\lambda} = (\Delta^0,0)\,,
\eeq
where $\Delta^0$ is a generic function of scalar invariants in the interaction. Therefore, in the forward scattering limit, the four-vector $N^{\lambda}$ becomes
\bea
N^{\lambda}&=&\epsilon^{\lambda\mu0\rho}  Q_{\mu}\Delta^{0}P_{\rho}
\nonumber\\&=& 4 \epsilon^{\lambda\mu0\rho}q_{\mu}\Delta^{0}p_{\rho}
\,,
\eea
and $\hat{e}^{(1)}$ is now defined as
\beq \label{e1}
\hat{e}^{(1) i}=\frac{N^{i}}{\sqrt{-\mathbf{N}^2}}~\quad \hat{e}^{(1) 0}= 0 \, ,
\eeq
where $\mathbf{N}^2$ will stand for the modulus square of the three-vector
\beq
N^{i}= 4\epsilon^{ij0k}q_{j}\Delta^{0}p_{k}~,
\eeq
which gives
\beq
\mathbf{N}^2=16(\Delta^{0})^2|\mathbf{p}\times\mathbf{q}|^2~.
\eeq
It is easy to verify that $F_{\mu \nu}$, with the new definitions of $\hat{e}^{(1)\lambda}$ and $\hat{e}^{(2)\lambda}$ in Eqs. \eqref{e1} and \eqref{e2}, satisfies the conserved current condition.

Before proceeding, it is worth to rewrite the factor of $G_2$ in Eq. \eqref{amplitude1} in a new form for the case of forward scattering. Using the following identity regarding the Levi-Civita pseudotensor \cite{Nieves:1982bq}
\beq
g_{\lambda\mu}\epsilon_{\nu\alpha\beta\gamma}-g_{\lambda\nu}\epsilon_{\mu\alpha\beta\gamma}+g_{\lambda\alpha}\epsilon_{\mu\nu\beta\gamma}
-g_{\lambda\beta}\epsilon_{\mu\nu\alpha\gamma}+g_{\lambda\gamma}\epsilon_{\mu\nu\alpha\beta}=0 \, ,
\eeq
we obtain
\bea\label{G2_term_verification}
\hat{e}^{(1) \mu}\hat{e}^{(2)\nu}-\hat{e}^{(2) \mu}\hat{e}^{(1) \nu}&=& \frac{4}{\sqrt{  m^2_f\mathbf{N}^2}}q_{\lambda}\Delta_{\beta}q_{\alpha}p_{\gamma}\left(g^{\nu\lambda}\epsilon^{\mu\alpha \beta \gamma}
-g^{\mu\lambda}\epsilon^{\nu\alpha \beta \gamma}\right)
\nonumber \\ &
=&\frac{4}{\sqrt{  m^2_f\mathbf{N}^2}}q_{\lambda}\Delta_{\beta}q_{\alpha}p_{\gamma}\left(g^{\lambda\alpha}\epsilon^{\mu\nu \beta\gamma}
-g^{\lambda \beta}\epsilon^{\mu\nu\alpha\gamma}+g^{\lambda\gamma}\epsilon^{\mu\nu\alpha \beta}\right)
\nonumber \\ &
=& \frac{4}{\sqrt{  m^2_f\mathbf{N}^2}}\left(q^2 \Delta_{\beta} p_{\gamma}\epsilon^{\mu\nu \beta\gamma}
-q\cdot\Delta\, q_{\alpha}p_{\gamma}\epsilon^{\mu\nu\alpha\gamma}+q\cdot p \, q_{\alpha}\Delta_{\beta}\epsilon^{\mu\nu\alpha \beta}\right) \, .
\eea
Thus, in the end we have
\beq \label{G2_term}
\hat{e}^{(1) \mu}\hat{e}^{(2)\nu}-\hat{e}^{(2) \mu}\hat{e}^{(1) \nu}= \frac{4}{\sqrt{  m^2_f\mathbf{N}^2}} \left(q^2 \Delta_{\alpha} p_{\beta}
-q\cdot\Delta\, q_{\alpha}p_{\beta} +q\cdot p \, q_{\alpha}\Delta_{\beta}\right)\epsilon^{\mu\nu\alpha \beta}   \, .
\eeq
Notice that the second term on the right-hand side of Eq. \eqref{G2_term} is equal in form to the one that appears at quantum level in the interaction of a photon with the magnetic moment of a neutrino (see Appendix \ref{loop}).  Using the definition of $\Delta^{\lambda}$ in Eq. \eqref{delta}, we can further simplify Eq. \eqref{G2_term} into\footnote{Here we are implicitly assuming that Greek indices take only Latin values. In fact, as we will see later on, only the Latin components of the Compton tensor will be important.}
\bea
\hat{e}^{(1) \mu}\hat{e}^{(2)\nu}-\hat{e}^{(2) \mu}\hat{e}^{(1) \nu}&=& \frac{4}{\sqrt{  m^2_f\mathbf{N}^2}} \left[(q_0^2 - q^2) \Delta^{0} \, p_{k}
+ (q\cdot p - q_0 \, p_0) \Delta^{0} \, q_k \right] \, \epsilon^{\mu\nu k 0}    \\ 
&=& \frac{4  \Delta^{0}}{\sqrt{  m^2_f\mathbf{N}^2}} \left[|\mathbf q|^2  \, p_{k}
-  (\mathbf{q} \cdot \mathbf{p})  \, q_k \right] \, \epsilon^{\mu\nu k 0}   \, .
\eea
Moreover, it is worth noticing that $\hat{e}^{(1)}$ is an axial vector and $\hat{e}^{(2)}$ is a vector. Using this property, it is straightforward to verify that the second and third brackets in Eq. \eqref{amplitude1} change sign under parity transformation, while the first and fourth brackets remain unchanged.
Both these two combinations of the Compton tensor satisfy the crossing symmetry and gauge-invariance. However, $F^{\mu\nu}$ can be even or odd under parity.

In order to discuss these cases, we first provide the general expression for the coefficients $G_i$, and then we start from the parity-invariant case by deriving all nonvanishing terms of each $G_i$ under the parity-invariance condition of the scattering amplitude. The coefficients can be represented in terms of the following bilinear covariant terms
\cite{Prange,Drechsel:2002ar,Lifshitz}
\beq
G_0=\bar{u}_{r'}\left[f_1+f_2\slashed{P}+f_3\gamma^{5}+f_4\gamma^{5}\slashed{P}\right]u_{r}\, ,\label{G00}
\eeq

\beq
G_1=\bar{u}_{r'}\left[f_5+f_6\slashed{P}+f_7\gamma^{5}+f_8\gamma^{5}\slashed{P}\right]u_{r}\, ,\label{G01}
\eeq

\beq
G_2=\bar{u}_{r'}\left[f_9+f_{10}\slashed{P}+f_{11}\gamma^{5}+f_{12}\gamma^{5}\slashed{P}\right]u_{r}\, ,\label{G02}
\eeq

\beq
G_3=\bar{u}_{r'}\left[f_{13}+f_{14}\slashed{P}+f_{15}\gamma^{5}+f_{16}\gamma^{5}\slashed{P}\right]u_{r}\, ,\label{G03}
\eeq
where  $u_r$ and $\bar{u}_{r'}$ are Dirac spinors associated to the fermion; $r,r'$ label fermion spin, $\slashed{P} =P_{\mu}\gamma^{\mu}$, $\gamma^{\mu}$, and $\gamma^{5}$ are Dirac matrices and $f_i$ are constant coefficients.

The invariant functions $G_i$ involve four possibilities. One can show that the $\slashed{Q}$ and $\slashed{t}$ terms are nothing else than numbers due to the Dirac equation and hence they do not appear in the $G_i$ invariants. Similarly, all higher powers of the $\gamma^\mu$ matrices are reduced to the above four possibilities. With the above representation, the time-reversal and parity transformations of each bilinear term are evident.

%%%%%%%%%%%%%%%%%%%%%%%%%%%%%%%%%%%%%%%%%%%%%%%%%%%%%%%%%%%%%%%%%%%%%%%%%%%%%%%%%%%%%%%%%%%%%%%%%%%%%%%%%%%%%%%%%%%%%%%%

\subsection{Even-parity amplitude}
In this subsection, we determine the form of the fermion-photon scattering amplitude with the condition that the amplitude is even under parity transformation. As we have seen, the photon scattering amplitude is represented by \eqref{amplitude0}.
Since under parity transformation, the polarization vectors change as
\beq
(\epsilon_0,\bm{\epsilon})\leftrightarrow (\epsilon_0,-\bm{\epsilon} )\, ,
\eeq
 the condition of parity invariance of scattering amplitude $M_{fi}$ implies
 \beq
 (F^{00},F^{i0},F^{ik})\rightarrow (F^{00},-F^{i0},F^{ik})\, .
 \eeq
Using the fact that $\hat{e}^{(1)}$ and $\hat{e}^{(2)}$ are a pseudovector and a vector, respectively, $G_0$ and $G_3$ must be scalars and  $G_1$ and $G_2$ must be pseudoscalars. Consequently, we can obtain the following constraints, as a result of the even-parity condition
 \beq
 f_3=f_4=f_5=f_6=f_9=f_{10}=f_{15}=f_{16}=0\, . \label{Ceven}
 \eeq
Then, we impose the condition of time-reversal invariance.
Under time-reversal we have
\beq
(q_0,\mathbf{q})\leftrightarrow (q'_0,-\mathbf{q}')~,~~~~~~~~~~(p_0,\mathbf{p})\leftrightarrow (p'_0,-\mathbf{p}')\, , \label{Tkp}
\eeq
and
\beq
(\epsilon_0,\bm{\epsilon})\leftrightarrow (\epsilon'^{\ast}_0,-\bm{\epsilon}'^{\ast} )\, .
\eeq
Hence, invariance of the scattering amplitude $M_{fi}$ under time-reversal yields
\beq
(F^{00},F^{i0},F^{ik})\rightarrow (F^{00},-F^{0i},F^{ki})\, .
\eeq
Similarly, the relations in Eq. \eqref{Tkp} imply
\bea
&&(Q_0,\mathbf{Q})\rightarrow (Q_0,-\mathbf{Q})\, ,~~~~~~~~~~~~~~~~(t_0,\mathbf{t})\rightarrow (-t_0, \mathbf{t})\, ,\nonumber \\ &&
(P_0,\mathbf{P})\rightarrow (P_0,-\mathbf{P})\, ,~~~~~~~~~~~~~~~~(N_0,\mathbf{N})\rightarrow (N_0, -\mathbf{N})\, ,
\eea
so that
\beq
\left(\hat{e}_0^{(1,2)},\hat{\bm{e}}^{(1,2)}\right)\rightarrow \left(\hat{e}_0^{(1,2)},-\hat{\bm{e}}^{(1,2)}\right)\, .
\eeq
Thus, invariance under time-reversal implies
\beq
G_{0,1,3}\rightarrow G_{0,1,3}~,~~~~~~~~~~~~~~ G_2\rightarrow -G_2\, ,
\eeq
and based on the following properties of spinor bilinear terms under a time-reversal transformation
\beq
\bar{u}'\gamma^{5}u\rightarrow -\bar{u}'\gamma^{5}u~,~~~~~~~~~~~~~~~~~~\bar{u}'\gamma^{5}\slashed{P}u\rightarrow  \bar{u}'\gamma^{5}\slashed{P}u\, ,
\eeq
one can verify the following additional conditions
\beq
f_7=f_{12}=0\, .
\eeq
Consequently, under parity and time reversal invariance the number of free coefficients is reduced to
\bea
&&G_0=\bar{u}_{r'}\left[f_1+f_2\slashed{P}\right]u_r\, ,~~~~~~~~~~~~
G_1=\bar{u}_{r'}f_8\gamma^{5}\slashed{P}u_r\, ,\nonumber \\ &&
G_2=\bar{u}_{r'}f_{11}\gamma^{5}u_r \, ,~~~~~~~~
G_3=\bar{u}_{r'}\left[f_{13}+f_{14}\slashed{P}\right]u_r \, . \label{G0i}
\eea
For further investigation, we analyze the transformation under charge conjugation and crossing. The charge conjugation leads to
\beq
(\epsilon_0,\bm{\epsilon})\leftrightarrow - (\epsilon^{\ast}_0,\bm{\epsilon}^{\ast})\, .
\eeq
As a result, invariance of the scattering amplitude $M_{fi}$ under C transformation leads to
\beq
(F^{00},F^{i0},F^{ik})\rightarrow (F^{00},F^{0i},F^{ki})\, .
\eeq
On the other hand, the crossing leads to
\beq
p \leftrightarrow -p'   ~~~~~~~~~~~\textrm{and} ~~~~~~~~~~~ \mu \leftrightarrow \nu \, ,
\eeq
and
\beq
\hat{e}^{(1) \lambda} \leftrightarrow \hat{e}^{(1) \lambda}  ~~~~~~~~ \hat{e}^{(2) \lambda} \leftrightarrow -\hat{e}^{(2) \lambda} \, ,
\eeq
and we find that under charge conjugation and crossing
\beq
G_{0,2,3} \rightarrow G_{0,2,3} ~~~~~~~~~  G_{1} \rightarrow -G_{1}\, ,
\eeq
that is satisfied by the results presented in \eqref{G0i}. Therefore, we can claim that the amplitude will be invariant under CPT and crossing symmetry.

In particular, let us discuss the standard Compton scattering amplitude, which is based on QED.
Using the standard Feynman rules, the amplitude of Compton scattering is given by
\beq
M_{fi}=-e^{2}\epsilon_{\mu}^s(p)~ \epsilon^{s' \ast}_{\nu}(p')[\bar{u}(q')Q^{\mu\nu}u(q)]\, ,
\eeq
where
\beq
Q^{\mu\nu}=\frac{1}{s-m^2}\gamma^{\nu}(\slashed p+ \slashed q+m)\gamma^{\mu}+\frac{1}{u-m^2}\gamma^{\mu}(\slashed q-\slashed p'+m)\gamma^{\nu}\, ,
\eeq
and the kinematic invariants are
\bea
s&=&(p+q)^2=(p'+q')^2=m^2+2p \cdot q=m^2+2p' \cdot q'\, ,\nonumber   \\
u&=&(p-q')^2=(p'-q)^2=m^2-2p \cdot q'=m^2-2p' \cdot q\, .
\eea
After some straightforward algebra we can find the following values of the coefficients $f_i$'s \cite{Lifshitz} 
\beq \label{QED}
\,f_1=-ma_{+}\, ,~~~~~~~~\,f_2=0\, ,~~~~~~~~f_8=\frac{1}{2}i a_{+}\, ,~~~~~~~~f_{11}=-ma_{+}\, ,~~~~~~~~f_{13}=m a_{+}\, ,~~~~~~~~f_{14}=\frac{1}{2} a_{-}\, ,
\eeq
where
\beq
a_{\pm}=\frac{1}{s-m^2} \pm \frac{1}{u-m^2}\, .
\eeq

\subsection{Odd-parity amplitude}

 In this subsection, we impose the odd-parity condition.
 In this case $F^{\mu\nu}$ is a pseudotensor that under parity operation must transform as
 \beq
 (F^{00},F^{i0},F^{ij})\rightarrow -(F^{00},-F^{i0},F^{ij})\, .
 \eeq
 Imposing the odd-parity condition and using the properties of bilinear terms under parity transformation we get
 \beq
 f_{1}=f_{2}=f_{7}=f_{8}=f_{11}=f_{12}=f_{13}=f_{14}=0\, .
 \eeq
 Therefore, those terms that remain after imposing the above condition are
 \bea
&&G_0=\bar{u}_{r'}\left[f_3\gamma^{5}+f_4\gamma^{5}\slashed{P}\right]u_r\, ,\:\:\:\:\:\:
G_1=\bar{u}_{r'}\left[f_5+f_6\slashed{P}\right]u_r\, ,\nonumber\\&&
G_2=\bar{u}_{r'}\left[f_9+f_{10}\slashed{P}\right]u_r\, ,\:\:\:\:\:\:\:\:\:\:\:\:\:\:
G_3=\bar{u}_{r'}\left[f_{15}\gamma^{5}+f_{16}\gamma^{5}\slashed{P}\right]u_r\, .\label{oddparity}
\eea
Afterward, imposing the even-time-reversal condition, we find
 \beq
 f_{3}=f_{9}=f_{10}=f_{15}=0\, .
 \eeq
 Thus, we remain with
 \bea
G_0=\bar{u}_{r'}f_4\gamma^{5}\slashed{P}u_r\, ,\:\:\:\:\:\:\:
G_1=\bar{u}_{r'}\left[f_5+f_6\slashed{P}\right]u_r\, ,\:\:\:\:\:\:\:
G_2=0\, ,\:\:\:\:\:\:\:
G_3=\bar{u}_{r'}f_{16}\gamma^{5}\slashed{P}u_r\, .\label{Gcoefficientsodd}
\eea
One can show that the resulting amplitude will be odd under charge conjugation. Therefore, the final form of the amplitude is even under CPT transformation.
In this case the $F^{\mu\nu}$ tensor is determined in terms of four free parameters. It is possible to compare our amplitudes with those of Kim and Dass in Ref. \cite{Kim:1975xf}.
Our results are consistent with the calculation of Kim and Dass which can be found also in Appendix \ref{A} .

%%%%%%%%%%%%%%%%%%%%%%%%%%%%%%%%%%%%%%%%%%%%%%%%%%%%%%%%%%%%%%%%%%%%%%%%%%%%%%%%%%%%%%%%%%%%%%%%%%%%%%%%%%%%%%%%%%%%%%%%
%%%%%%%%%%%%%%%%%%%%%%%%%%%%%%%%%%%%%%%%%%%%%%%%%%%%%%%%%%%%%%%%%%%%%%%%%%%%%%%%%%%%%%%%%%%%%%%%%%%%%%%%%%%%%%%%%%%%%%%%

\section{Forward Scattering Term} \label{4}

In this section we will provide a general expression for the forward scattering term on the right-hand side of the Boltzmann equation \eqref{Boltzmann equation}.

The general form of the interaction Hamiltonian defined in \eqref{intH2} can be written as \cite{Kosowsky:1994cy}
 \bea
 H_I(t)&=&\int d\mathbf{q}d\mathbf{q}'d\mathbf{p}d\mathbf{p}'(2\pi)^{3}\delta^{(3)}(\mathbf{q}'+\mathbf{p}'-\mathbf{q}-\mathbf{p})
 \exp\left[it\left(q'^{0}+p'^{0}-q^{0}-p^{0}\right)\right]\nonumber \\&&
 \times \left[b^{\dag}_{r'}(q')a^{\dag}_{s'}(p')\bar{u}_{r'}(q')F^{\mu\nu}(qr,q'r',ps,p's')u_{r}(q)\epsilon^{s}_{\mu}(\mathbf{p})\epsilon^{s'}_{\nu}(\mathbf{p}')a_{s}(p)b_{r}(q)\right]\, ,
 \label{intH22}
 \eea
where
\beq
d\mathbf{q}=\frac{d^{3}\mathbf{q}}{(2\pi)^{3}}\frac{m_f}{q^{0}}\, , \:\:\:\:\;\:\:\:\:\:\:\:\:\:\;\:\:\:\:\:
d\mathbf{p}=\frac{d^{3}\mathbf{p}}{(2\pi)^{3}2p^{0}}\, .
\eeq
 $a_{s}$ and $a^{\dag}_{s'}$ are photon annihilation and creation operators, respectively, which satisfy
the canonical commutation relations
\beq
\left[a_s(\mathbf{p}),a^{\dag}_{s'}(\mathbf{p}')\right]=(2\pi)^32p^0\delta^{(3)}(\mathbf{p}-\mathbf{p}')\delta_{ss'}\, ,
\eeq
and $b^{(r)}$ and $b^{(r)\,\dag}$ are fermion annihilation and creation operators, respectively, obeying
the canonical anti-commutation relations
\beq
\left\{b_r(\mathbf{q}),b^{\dag}_{r'}(\mathbf{q}')\right\}=(2\pi)^3\frac{q^0}{m_{f}}\delta^{(3)}(\mathbf{q}-\mathbf{q}')\delta_{rr'}\, ,
\eeq
where $m_f$ is the fermion mass.

Using Eq. \eqref{intH22}, the commutation relation in the forward scattering term of Eq. \eqref{Boltzmann equation} becomes
\beq \begin{split} \label{forward}
\left[H_I
(0),\mathcal{D}_{ij}(k)\right]&=\int d\mathbf{q}d\mathbf{q}'d\mathbf{p}d\mathbf{p}'(2\pi)^{3}\delta^{(3)}(\mathbf{q}'+\mathbf{p}'-\mathbf{q}-\mathbf{p})
\bar{u}_{r'}(q')F^{\mu\nu}(qr,q'r',ps,p's')u_{r}(q)\epsilon^{s}_{\mu}(\mathbf{p})\epsilon^{s'}_{\nu}(\mathbf{p}') \\
&\times \left[b^{\dag}_{r'}(q')b_{r}(q)a^{\dag}_{s'}(p')a_{j}(k)2p^{0}(2\pi)^{3}\delta_{is}\delta^{(3)}(\mathbf{p}-\mathbf{k})
  -b^{\dag}_{r'}(q')b_{r}(q)a^{\dag}_{i}(k)a_{s}(p)2p'^{0}(2\pi)^{3}\delta_{js'}\delta^{(3)}(\mathbf{p}'-\mathbf{k})\right]\, .
\end{split}\eeq
After this step, in order to evaluate the forward scattering term, we will need to take the expectation value of Eq. \eqref{forward}. For this purpose, we provide the following expectation values \cite{Kosowsky:1994cy}:
\beq
\left<a^{\dag}_{m}(p')a_{n}(p)\right>=2p^0(2\pi)^3\delta^{(3)}(\mathbf{p}-\mathbf{p}')\rho_{mn}(\mathbf{p})\, ,
\eeq
and
\beq
\left<b^{\dag}_{m}(q')b_{n}(q)\right>=\frac{q^0}{m_f}(2\pi)^3\delta^{(3)}(\mathbf{q}-\mathbf{q}')\delta_{mn}\frac{1}{2}n_{f}(\mathbf{q})\, ,
\eeq
where $\rho_{mn}$ is the photon beam polarization matrix and $n_{f}$ is the number density of fermions of momentum $\mathbf{q}$ per unit volume. After using the Dirac delta functions, one can easily perform the integrations over $\mathbf{p}$,
$\mathbf{p}'$, and $\mathbf{q}'$ and obtain the limit $p=p'$ and $q=q'$ of the integrand, in agreement with the forward scattering condition.

At this point, we can fix the Coulomb gauge for the photon polarization vectors, where we have $\epsilon^{\mu}=(0,\bm{\epsilon})$. As a consequence of this gauge-fixing, we are interested in only ``Latin" components of the Compton tensor $F^{\mu \nu}$ (thus, Latin components of the vector bases $\hat e^{(1)}$ and $\hat e^{(2)}$) to do the contractions in Eq. \eqref{forward}. In particular,  using the definitions \eqref{e1} and \eqref{e2} and the result \eqref{G2_term}, the $F^{ij}$ components in the forward scattering limit can be represented as
\beq \begin{split}
\bar{u}_{r'}(q') F^{ij} u_r(q)&=\left(G_0+G_3\right)\hat{e}^{(1)i}\hat{e}^{(1)j}+\left(G_0-G_3\right)\hat{e}^{(2)i}\hat{e}^{(2)j}
+G_1\left(\hat{e}^{(1) i}\hat{e}^{(2) j}+\hat{e}^{(2) i}\hat{e}^{(1) j}\right)
+G_2\left(\hat{e}^{(1) i}\hat{e}^{(2) j}-\hat{e}^{(2) i}\hat{e}^{(1) j}\right) \\
&=\left(G_0+G_3\right)\frac{(4\Delta^0)^2}{\mathbf{N}^2}(\mathbf{q}\times\mathbf{p})^i(\mathbf{q}\times\mathbf{p})^j+\left(G_0-G_3\right)\frac{q^i}{m_f}\frac{q^j}{m_f}
+ G_1\frac{4\Delta^0}{\sqrt{m^2_f\mathbf{N}^2}}\left[(\mathbf{q}\times\mathbf{p})^iq^j+q^i(\mathbf{q}\times\mathbf{p})^j\right]\\
&+\,G_2\frac{4 \Delta^0}{\sqrt{  m^2_f\mathbf{N}^2}} \left[|\mathbf q|^2  \, p_{k}
-  (\mathbf{q} \cdot \mathbf{p})  \, q_k \right] \, \epsilon^{i j k 0}\, .\label{Fijtensor}
\end{split}\eeq

In the next sections, we will study the phenomenological consequences for CMB polarization of the forward scattering term in specific cases.

%%%%%%%%%%%%%%%%%%%%%%%%%%%%%%%%%%%%%%%%%%%%%%%%%%%%%%%%%%%%%%%%%%%%%%%%%%%%%%%%%%%%%%%%%%%%%%%%%%%%%%%%%%%%%%%%%%%%
%%%%%%%%%%%%%%%%%%%%%%%%%%%%%%%%%%%%%%%%%%%%%%%%%%%%%%%%%%%%%%%%%%%%%%%%%%%%%%%%%%%%%%%%%%%%%%%%%%%%%%%%%%%%%%%%%%%%

\section{General conditions for generating circular polarization} \label{5}

In this section, we will give the most general conditions for generating circular polarization from photon-fermion forward scattering. Thus, we will consider specific expressions of the Compton tensor \eqref{Fijtensor}, evaluate Eq. \eqref{forward}, and study the effects of new interactions on the Stokes parameters.

%%%%%%%%%%%%%%%%%%%%%%%%%%%%%%%%%%%%%%%%%%%%%%%%%%%%%%%%%%%%%%%%%%%%%%%%%%%%%%%%%%%%%%%%%%%%%%%%%%%%%%%%%%%%%%%%%%%%

\subsection{ Even-parity amplitude }

We start by considering the even-parity terms.
The general forms of the $G_i$ coefficients invariant under time-reversal have been derived in the previous section. We have also determined the coefficients for the QED case.

The coefficients $G_i$ read
\bea
&&G_0+G_3=\bar{u}_{r'}(\tilde{f}_1+\tilde{f}_2 \slashed{P})u_r\, ,~~~~~~~~~~~~
G_1=\bar{u}_{r'}(\tilde{f}_3\gamma^{5} \slashed{P})u_r\, ,\nonumber \\ &&
G_2=\bar{u}_{r'}(\tilde{f}_{4}\gamma^{5} )u_r\, ,~~~~~~~~~~~~~~~~~~~~~~~~~
G_0-G_3=\bar{u}_{r'}(\tilde{f}_{5}+\tilde{f}_{6} \slashed{P})u_r\, .
\eea
where $\tilde{f}_1=f_1+f_{13}$, $\tilde{f}_2=f_2+f_{14}$, $\tilde{f}_3=f_8$, $\tilde{f}_4=f_{11}$, $\tilde{f}_5=f_1-f_{13}$ and $\tilde{f}_6=f_2-f_{14}$.

Using the well-known spinorial relations
\beq
\bar{u}_{r'}(q)\gamma^{5}u_{r}(q)= 0\, ,
\eeq
and
\beq
\bar{u}_{r'}(q)\gamma^{\mu}u_{r}(q)=\delta_{rr'}\frac{q^{\mu}}{m_f}\, ,
\eeq
we find
\bea
&&G_0+G_3= \left(\tilde{f}_1+ \tilde{f}_2 \frac{P\cdot q}{m_f}\right) \delta_{rr'}\, ,~~~
G_1=\tilde{f}_3 \bar{u}_{r'} \gamma^5\slashed{P}  u_r\, ,~~~
G_2= 0\, ,~~~
G_0-G_3=\left(\tilde{f}_5+ \tilde{f}_6 \frac{P\cdot q}{m_f}\right) \delta_{rr'}\, .
\eea
Using these results, the scattering amplitude is simplified considerably to
\bea
M_{fi}&=&\left( \tilde{f}_1+ \tilde{f}_2 \frac{P\cdot q}{m_f} \right)\frac{(4\Delta^0)^2}{\mathbf{N}^2}(\mathbf{q}\times\mathbf{p})\cdot \bm{\epsilon}^{s}(\mathbf{q}\times\mathbf{p})\cdot \bm{\epsilon}^{s'}\delta_{rr'}+ \left( \tilde{f}_5+ \tilde{f}_6 \frac{P\cdot q}{m_f} \right)\frac{(\mathbf{q}\cdot \bm{\epsilon}^{s})}{m_f}\frac{(\mathbf{q}\cdot \bm{\epsilon}^{s'})}{m_f}\delta_{rr'}
\nonumber \\ &&\!\!\!\!
+\tilde{f}_3 \bar{u}_{r'} \gamma^5 \slashed{P} u_r \frac{4\Delta^0}{\sqrt{m^2_f\mathbf{N}^2}}\left[(\mathbf{q}\times\mathbf{p})\cdot \bm{\epsilon}^{s}\, (\mathbf{q}\cdot \bm{\epsilon}^{s'})
+(\mathbf{q}\cdot \bm{\epsilon}^{s})(\mathbf{q}\times\mathbf{p})\cdot \bm{\epsilon}^{s'}\right]
\, .\label{amplitude3}
\eea
In this equation the main effects are expected to come from the term multiplying the $\tilde{f}_5$ and $\tilde{f}_6$. In fact, other terms, containing at least one factor of $\Delta^0$, will appear only when considering loop quantum effects. For this reason, in the next steps we will ignore them, since in a perturbation quantum field theory framework they are supposed to be an higher-order effect.
Thus, the time evolution of polarization matrix elements is given by (from now on we will explicitly account for
spatial dependence in the Boltzmann equations)
\bea
	 \frac{d} {dt} \rho_{ij}({\bf{ x, k}})&=& \frac{i}{2k^{0}m^2_f} \int d\mathbf{q}\,n_f(\mathbf{x, q})\,\left( \tilde{f}_5+2 \tilde{f}_6 \frac{q\cdot k}{m_f} \right)\,\left(\delta_{is}\rho_{s'j}(\mathbf{x, k})-\delta_{js'}\rho_{is}(\mathbf{x, k})\right)\,\left(\mathbf{q}\cdot \bm{\epsilon}^{s}\right)\,(\mathbf{q}\cdot \bm{\epsilon}^{s'}) \nonumber \\ &&\!\!\!\!
\:\:\:\:\:\:\:\:\:\:\:\: + \textrm{standard Compton scattering terms (s.C.s.t.)}\, . \label{Stokesevolv1}
\eea
Now, expressing Eq. \eqref{Stokesevolv1} in terms of the different components, we have
\bea
   \frac{d}{dt}\, \rho^{(1)}_{11}(\mathbf{x, k})&=& \frac{i}{2k^{0}m^2_f} \int d\mathbf{q}\,n_{f}(\mathbf{x, q})
     \,\left( \tilde{f}_5+2 \tilde{f}_6 \frac{q\cdot k}{m_f}\right)\, (\mathbf{q}\cdot \bm{\epsilon}_{2})\,(\mathbf{q} \cdot \bm{\epsilon}_{1})\left[\rho^{(1)}_{21}(\mathbf{x, k})-\rho^{(1)}_{12}(\mathbf{x, k})\right]
      \texttt{\texttt{}}   +\textrm{s.C.s.t.}\, ,
\eea
  \bea
 \frac{d}{dt}\,\rho^{(1)}_{22}(\mathbf{x, k})&=& -\frac{d}{dt}\, \rho^{(1)}_{11}(\mathbf{x, k})\, ,
\eea
\bea
   \frac{d}{dt}\, \rho^{(1)}_{12}(\mathbf{x, k})&=&\frac{i}{2k^{0}m^2_f}\int d\mathbf{q}\,n_{f}(\mathbf{x, q})\,\left( \tilde{f}_5+2 \tilde{f}_6 \frac{q\cdot k}{m_f} \right)\,
 \left[(\mathbf{q}\cdot \bm{\epsilon}_{2})\,(\mathbf{q}\cdot \bm{\epsilon}_{1})(\rho^{(1)}_{22}(\mathbf{x, k})-\rho^{(1)}_{11}(\mathbf{x, k}))+\right. \nonumber\\
&&\left. [(\mathbf{q}\cdot \bm{\epsilon}_{1})^2-(\mathbf{q}\cdot \bm{\epsilon}_{2})^2]\rho^{(1)}_{12}(\mathbf{x, k})\right]+  \textrm{s.C.s.t.}\, ,
\eea
\bea
   \frac{d}{dt}\, \rho^{(1)}_{21}(\mathbf{x, k})&=&-\frac{i}{2k^{0}m^2_f}\int d\mathbf{q}\,n_{f}(\mathbf{x, q})\,\left( \tilde{f}_5+2 \tilde{f}_6 \frac{q\cdot k}{m_f} \right)\,
 \left[(\mathbf{q}\cdot \bm{\epsilon}_{2})\,(\mathbf{q}\cdot \bm{\epsilon}_{1})(\rho^{(1)}_{22}(\mathbf{x, k})-\rho^{(1)}_{11}(\mathbf{x, k}))\right.\nonumber\\
&&\left.+[(\mathbf{q}\cdot \bm{\epsilon}_{1})^2-(\mathbf{q}\cdot \bm{\epsilon}_{2})^2]\rho^{(1)}_{21}(\mathbf{x, k})\right]+ \textrm{s.C.s.t.}\, .
\eea
We can also convert the density matrix elements to the normalized Stokes brightness perturbations after changing momentum to the comoving one, $k_c=ak$, and going to the Fourier space. We find
 \bea
  \frac{d}{d\eta}\, {I}^{(S)}(\mathbf{K, \mathbf{k}_c})=\textrm{s.C.s.t.}\, ,
\eea
 \bea
  \frac{d}{d\eta}\, {Q}^{(S)}(\mathbf{K, \mathbf{k}_c})&=&-\frac{ a^2(\eta)}{  k_c^0 m^2_f}\,\int d\mathbf{q}\,n_{f}(\mathbf{K, q})\,\left( \tilde{f}_5+2 \tilde{f}_6 \frac{q\cdot k_c}{a(\eta) m_f} \right)
      \, (\mathbf{q}\cdot \bm{\epsilon_{2}})\,(\mathbf{q}\cdot \bm{\epsilon}_{1} ) {V}^{(S)}(\mathbf{K, k_c})
       +\textrm{s.C.s.t.}\, ,
\eea
\bea
  \frac{d}{d\eta}\, {U}^{(S)}(\mathbf{K, \mathbf{k}_c})=
\frac{ a^2(\eta)}{ 2  k_c^0 m^2_f}\,\int d\mathbf{q}\,n_{f}(\mathbf{K, q})\,\left( \tilde{f}_5+2 \tilde{f}_6 \frac{q\cdot k_c}{a(\eta) m_f} \right)\,
     [(\mathbf{q}\cdot \bm{\epsilon}_{1})^2-(\mathbf{q}\cdot \bm{\epsilon}_{2})^2] {V}^{(S)}(\mathbf{K, k_c})
       +\textrm{s.C.s.t.}\, ,
\eea
\bea
 \frac{d}{d \eta} {V}^{(S)}(\mathbf{K, \mathbf{k}_c}) &=& -\frac{a^2(\eta)}{2 k_c^0 m^2_f}\int d\mathbf{q}\,n_{f}(\mathbf{K, q})\,\left( \tilde{f}_5+2 \tilde{f}_6 \frac{q\cdot k_c}{a(\eta) m_f} \right)\, \left[-2(\mathbf{q}\cdot \bm{\epsilon_{2}}) (\mathbf{q}\cdot \bm{\epsilon}_{1}) {Q}^{(S)}(\mathbf{K, k_c})\right.\nonumber\\
&&\left.+[(\mathbf{q}\cdot \bm{\epsilon}_{1})^2-(\mathbf{q}\cdot \bm{\epsilon}_{2})^2] {U}^{(S)}(\mathbf{K, k_c})\right]+\textrm{s.C.s.t.}\, .
\eea
From the last set of equations we see that the V-modes in the CMB can be generated even with a parity preserving interaction. In particular, it is straightforward to verify that the fermionic number density $n_{f}(\mathbf{K,q})$ has to contain anisotropies in order to achieve a nontrivial coupling. In fact, under the assumption that $n_{f}(\mathbf{K,q})$ does not contain anisotropies, using the generic parametrizations \eqref{par_momenta}  the angular integrals over the fermionic momentum $\mathbf q$ are vanishing as we show in the following:
\begin{align}
&\int_0^{\pi} d\theta' \sin \theta'\,\int_0^{2 \pi} d\varphi' \, \, \left( \tilde{f}_5+2 \tilde{f}_6 \frac{q\cdot k_c}{a(\eta) m_f} \right)\, (\mathbf{q}\cdot \bm{\epsilon}_{2})(\mathbf{q}\cdot \bm{\epsilon}_{1}) \propto \int_0^{\pi} d\theta' \sin \theta'\,\int_0^{2 \pi} d\varphi' \,\nonumber \\
&\times \left( \tilde{f}_5+2 \tilde{f}_6 \frac{\cos\theta \cos\theta' + \cos(\varphi - \varphi') \sin\theta \sin\theta'}{a(\eta) m_f} \right)\, \sin(\varphi - \varphi') \sin\theta' \, \left[\cos\theta' \sin\theta - \cos(\varphi - \varphi') \cos\theta \sin\theta'\right]= 0 \, ,
\end{align}
and
\begin{align}
&\int_0^{\pi} d\theta' \sin \theta'\,\int_0^{2 \pi} d\varphi' \,\, \left( \tilde{f}_5+2 \tilde{f}_6 \frac{q\cdot k_c}{a(\eta) m_f} \right)\, [(\mathbf{q}\cdot \bm{\epsilon}_{1})^2-(\mathbf{q}\cdot \bm{\epsilon}_{2})^2] \propto \int_0^{\pi} d\theta' \sin \theta'\,\int_0^{2 \pi} d\varphi' \,\nonumber \\
&\times \left( \tilde{f}_5+2 \tilde{f}_6 \frac{\cos\theta \cos\theta' + \cos(\varphi - \varphi') \sin\theta \sin\theta'}{a(\eta) m_f} \right)\, \left[ \,(\cos\theta' \sin\theta - 
   \cos(\varphi - \varphi') \cos\theta \sin\theta')^2 -\sin^2(\varphi - \varphi') \sin^2\theta' \,\right] \nonumber \\
   & = 0 \, . 
\end{align}
Moreover, from the current model of particle physics, we know that a fermion can have a parity preserving interaction with a photon only through QED vertices. If we take the values of $\tilde{f}_{5}$ and $\tilde{f}_6$ for the case of QED, Eq. \eqref{QED}, and we evaluate them in the forward scattering limit, we find that $\tilde{f}_5 =\tilde{f}_6= 0$. Thus, QED does not provide mixing terms among different polarizations, and only a parity preserving theory which goes beyond the standard paradigm could provide some kind of V-mode generation.

%%%%%%%%%%%%%%%%%%%%%%%%%%%%%%%%%%%%%%%%%%%%%%%%%%%%%%%%%%%%%%%%%%%%%%%%%%%%%%%%%%%%%%%%%%%%%%%%%%%%%%%%%%%%%%%%%%%%%%%%

\subsection{Odd-parity amplitude} \label{odd_parity}

\begin{figure}[H]
	\centering
	\includegraphics[width=3in]{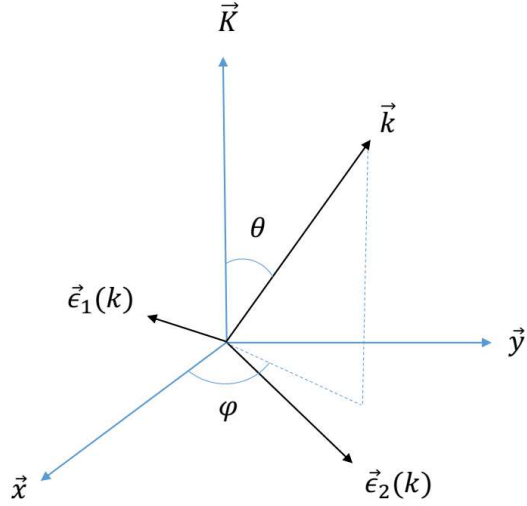}
	\caption{Pictorial representation of the polarizations and momentum direction of the photon.} \label{fig:third_sub}
\end{figure}

The general form of scattering amplitude for odd-parity was derived in Sec. \ref{3}. In that section, we found the general form of coefficients $G_i$ for the odd-parity case
 \bea
G_0=\bar{u}_{r'}(f_4\gamma^{5}\slashed{P})u_r\, ,\:\:\:\:\:\:\:
G_1=\bar{u}_{r'}(f_5+f_6\slashed{P})u_r\, ,\:\:\:\:\:\:\:
G_2=0\, ,\:\:\:\:\:\:\:
G_3=\bar{u}_{r'}(f_{16}\gamma^{5}\slashed{P})u_r\, .\label{Gcoefficients1}
\eea
The amplitude can be constructed using the tensor \eqref{Fijtensor} and replacing the values of the coefficients \eqref{Gcoefficients1}.
As in the previous subsection, we focus only on the terms which are expected to give the dominant contributions. Thus, our amplitude reads
\bea
M_{fi}=
(f_4-f_{16})\bar{u}_{r'}(q)\gamma^{5} \slashed{P}\,u_{r}(q) \,\frac{(\mathbf{q}\cdot \bm{\epsilon}^{s})}{m_f}\,\frac{(\mathbf{q}\cdot \bm{\epsilon}^{s'})}{m_f}
\, .
\eea
 Using this result, we can find the time evolution of polarization matrix elements as
\bea
	 \frac{d} {dt} \rho_{ij}({\bf {x, k}})&=&i\frac{f_{\textrm{p}}}{4k^{0}m^2_f}\int d\mathbf{q}\,n_f(\mathbf{x, q})\left(\delta_{is}\rho_{s'j}(\mathbf{x, k})-\delta_{js'}\rho_{is}(\mathbf{x, k})\right)\bar{u}_{r}(q)\gamma^{5}\slashed{k}\,u_{r}(q) \,(\mathbf{q}\cdot \bm{\epsilon}^{s})\,(\mathbf{q}\cdot \bm{\epsilon}^{s'})
+ \textrm{s.C.s.t.}\, , \label{Stokesevolv}
\eea
where $f_{\textrm{p}}\equiv2 (f_4-f_{16})$. Therefore, we have
 \bea
   \frac{d}{dt}\, \rho^{(1)}_{11}(\mathbf{x, k})&=&- \frac{if_{\textrm{p}}}{4k^{0}m^2_f}\int d\mathbf{q}\,n_{f}(\mathbf{x, q})
 \bar{u}_{r}k\!\!\!/\gamma^{5}u_{r}
     \, (\mathbf{q}\cdot \bm{\epsilon}_{2})\,(\mathbf{q}\cdot \bm{\epsilon}_{1})\left[\rho^{(1)}_{21}(\mathbf{x, k})-\rho^{(1)}_{12}(\mathbf{x, k})\right]
      \texttt{\texttt{}}   +\textrm{s.C.s.t.}\, ,
\eea
  \bea
 \frac{d}{dt}\,\rho^{(1)}_{22}(\mathbf{x, k})&=& -\frac{d}{dt}\, \rho^{(1)}_{11}(\mathbf{x, k})\, ,
\eea
\beq \begin{split}
   \frac{d}{dt}\, \rho^{(1)}_{12}(\mathbf{x, k})=&-\frac{if_{\textrm{p}}}{4k^{0}m^2_f}\int d\mathbf{q}\,n_{f}(\mathbf{x, q})
 \bar{u}_{r}k\!\!\!/\gamma^{5}u_{r}
 \left[(\mathbf{q}\cdot \bm{\epsilon}_{2})\,(\mathbf{q}\cdot \bm{\epsilon}_{1})(\rho^{(1)}_{22}(\mathbf{x, k})-\rho^{(1)}_{11}(\mathbf{x, k}))+[(\mathbf{q}\cdot \bm{\epsilon}_{1})^2-(\mathbf{q}\cdot \bm{\epsilon}_{2})^2]\rho^{(1)}_{12}(\mathbf{x, k})\right]\\
 &+ \,\, \textrm{s.C.s.t.}\, ,
\end{split}
\eeq

\beq\begin{split}
   \frac{d}{dt}\, \rho^{(1)}_{21}(\mathbf{x, k}) =& \frac{if_{\textrm{p}}}{4k^{0}m^2_f}\int d\mathbf{q}\,n_{f}(\mathbf{x, q})
 \bar{u}_{r}k\!\!\!/\gamma^{5}u_{r}
 \left[(\mathbf{q}\cdot \bm{\epsilon}_{2})\,(\mathbf{q}\cdot \bm{\epsilon}_{1})(\rho^{(1)}_{22}(\mathbf{x, k})-\rho^{(1)}_{11}(\mathbf{x, k}))+[(\mathbf{q}\cdot \bm{\epsilon}_{1})^2 - (\mathbf{q}\cdot \bm{\epsilon}_{2})^2]\rho^{(1)}_{21}(\mathbf{x, k})\right] \\
 &+ \,\, \textrm{s.C.s.t.}\, .
\end{split}
\eeq
Here, we convert the density matrix elements to the normalized Stokes brightness perturbations and go to the Fourier space to obtain
 \bea
  \frac{d}{d\eta}\, {I}^{(S)}(\mathbf{K, k_c})=\textrm{s.C.s.t.}\, ,
\eea
 \bea
  \frac{d}{d \eta}\, {Q}^{(S)}(\mathbf{K, k_c})&=&\frac{a(\eta)\,f_{\textrm{p}}}{ 2 k_c^0 m^2_f}\, \int d\mathbf{q}\,n_{f}(\mathbf{K, q})\,\bar{u}_r\slashed{k}_c \gamma^5 u_r
      \, (\mathbf{q}\cdot \bm{\epsilon_{2}})\,(\mathbf{q}\cdot \bm{\epsilon}_{1}) {V}^{(S)}(\mathbf{K, k_c})
       +\textrm{s.C.s.t.}\, ,\label{E-SPQ}
\eea
\bea
  \frac{d}{d \eta}\, {U}^{(S)}(\mathbf{K, k_c})=-
 \frac{a(\eta)\,f_{\textrm{p}}}{ 4 k_c^0 m^2_f}\,\int d\mathbf{q}\,n_{f}(\mathbf{K, q})\,\bar{u}_r\slashed{k}_c \gamma^5 u_r\,
     [(\mathbf{q}\cdot \bm{\epsilon}_{1})^2 - (\mathbf{q}\cdot \bm{\epsilon}_{2})^2] {V}^{(S)}(\mathbf{K, k_c})
       +\textrm{s.C.s.t.}\, ,\label{E-SPU}
\eea
\begin{align}
 \frac{d}{d \eta}\, {V}^{(S)}(\mathbf{K, k_c})=&\frac{a(\eta)\,f_{\textrm{p}}}{4  k_c^0 m^2_f}\, \int d\mathbf{q}\,n_{f}(\mathbf{K, q})\,\bar{u}_r\slashed{k}_c \gamma^5 u_r\,
 \left[-2(\mathbf{q}\cdot \bm{\epsilon_{2}})\,(\mathbf{q}\cdot \bm{\epsilon}_{1}) {Q}^{(S)}(\mathbf{K, k_c})+[(\mathbf{q}\cdot \bm{\epsilon}_{1})^2 - (\mathbf{q}\cdot \bm{\epsilon}_{2})^2] {U}^{(S)}(\mathbf{K, k_c})\right]\nonumber\\
&+ \textrm{s.C.s.t.} \, .~~~\label{E-SPV}
\end{align}
The quantity $\sum_r \bar{u}_r \gamma^\mu \gamma^5 u_r$ vanishes when we sum over spins if the interacting fermion exists in both left- or right-handed helicity states.
Thus, looking to this final set of equations, circular polarization in the CMB photons can be generated from a parity violating interaction only if the following condition is satisfied:
\beq
 \sum_r \bar{u}_r \gamma^\mu \gamma^5 u_r\neq 0\, ,
\eeq
implying that the fermion particle must interact only in left- or right-handed helicity state.\\\\
Now, in order to perform the integral over $\mathbf{q}$, we choose the momentum and photon polarization vectors in the following form (see Fig. \ref{fig:third_sub}):
\bea 
\hat{\mathbf{K}}&=&(0,0,1)\, , \nonumber\\
 \hat{\mathbf{k}}&=&(\sin\theta\cos\varphi,\,\sin\theta\sin\varphi,\,\cos\theta)\, , \nonumber \\
\hat{\mathbf{q}}&=&(\sin\theta'\cos\varphi',\,\sin\theta'\sin\varphi',\,\cos\theta')\, ,\nonumber \\
 \bm{\epsilon}_{1}(k)&=& (\cos\theta\cos\varphi,\,\cos\theta\sin\varphi,\,-\sin\theta)\, , \nonumber\\
 \bm{\epsilon}_{2}(k)&=& (-\sin\varphi,\,\cos\varphi,\,0)~. \label{par_momenta}
\eea
In particular, we can expand $n_f(\mathbf{K, q})$ as  \cite{Ma:1995, Weinberg:2008zzc}
\beq
n_f(\mathbf{K}, \mathbf{q})=n_f(\mathbf{K}, |\mathbf{q}|)\sum_{\ell,m}c_{\ell m}Y_{\ell}^{m}(\mathbf{\hat{q}})\, . \label{nexpand}
\eeq
The number density mediated over all the possible fermionic momenta is given by
\bea
\bar{n}_f(\mathbf{K})&=&
\sum_{\ell,m}\int \frac{d^{3}\mathbf{q}}{(2\pi)^{3}}\,n_f(\mathbf{K}, |\mathbf{q}|)\,c_{\ell m}Y_{\ell}^{m}(\mathbf{\hat{q}})
\nonumber \\&&
=\frac{c_{00}}{(2\pi)^{2}\sqrt{\pi}}\int d|\mathbf{q}||\mathbf{q}|^{2}\,n_f(\mathbf{K}, |\mathbf{q}|)\, .
\eea
Now, considering a left-handed fermion and including only the lowest multipole moments of expansion \eqref{nexpand}, the final mixing terms produced by the forward scattering term read as (for the complete equations, see Appendix \ref{app:C})
\beq\begin{split}
   \frac{d}{d\eta}\, {Q}^{(S)}(\mathbf{K, k_c})=-& i \sqrt{\frac{2\pi}{15}}\frac{a(\eta)\,f_{\textrm{p}}}{ 16\pi^3  m^2_f}\,\int d|\mathbf{q}| \frac{ |\mathbf{q}|^5}{q^0}n_f(\mathbf{K}, |\mathbf{q}|)  \left[\left(c_{2-2} e^{-2i\varphi}-c_{22} e^{2i\varphi}\right) \cos \theta-\left(c_{21} e^{i\varphi}+c_{2-1}e^{-i\varphi}\right) \sin \theta\right]\\
&\times \, {V}^{(S)}(\mathbf{K,k_c}) +\,\,\textrm{s.C.s.t.}\, ,
\end{split}
\eeq
\beq \begin{split}
 \frac{d}{d\eta}\, {U}^{(S)}(\mathbf{K, k_c})=&
\sqrt{\frac{2\pi}{15}}\frac{a(\eta)\,f_{\textrm{p}}}{ 32 \pi^3  m^2_f}\,\int d|\mathbf{q}| \frac{ |\mathbf{q}|^5}{q^0}n_f(\mathbf{K}, |\mathbf{q}|)   \,\left\{\sqrt{6}  \sin^2\theta\, c_{20}+ \left[\left(c_{21}e^{i\varphi}-c_{2-1} e^{-i\varphi}\right) \sin 2\theta \right. \right. \\
&  \left.\left.+\left(c_{22} e^{2i\varphi}+c_{2-2} e^{-2i\varphi}\right)(1+ \cos ^2\theta) \right]\right\}{V}^{(S)}(\mathbf{K, k_c})+\textrm{s.C.s.t.}\, ,
\end{split}
\eeq
\beq \begin{split}
\frac{d}{d\eta}\, {V}^{(S)}(\mathbf{K, k_c})&= \sqrt{\frac{2\pi}{15}}\frac{a(\eta)\,f_{\textrm{p}}}{ 32 \pi^3  m^2_f}\,\int d|\mathbf{q}| \frac{ |\mathbf{q}|^5}{q^0}  n_f(\mathbf{K}, |\mathbf{q}|) \,
\left\{2i\left[\left(c_{2-2} e^{-2i\varphi}-c_{22} e^{2i\varphi}\right) \cos \theta-\left(c_{21} e^{i\varphi}+c_{2-1}e^{-i\varphi}\right) \sin \theta\right]  \right.\\
&\left. \times \, {Q}^{(S)}(\mathbf{K, k_c}) -\left\{\sqrt{6}  \sin^2\theta\, c_{20}+  \left[\left(c_{21}e^{i\varphi}-c_{2-1} e^{-i\varphi}\right) \sin 2\theta +\left(c_{22} e^{2i\varphi}+c_{2-2} e^{-2i\varphi}\right)( 1+\cos^ 2\theta) \right]\right\} \right. \\
& \left. \, \times \, {U}^{(S)}(\mathbf{K, k_c})\right\}+\,\,\textrm{s.C.s.t.}\, .~~
\end{split}
\eeq
From this set of equations we find the following second condition for generating circular polarization: 
\beq
     \left(c_{2-2} e^{-2i\varphi}-c_{22} e^{2i\varphi}\right) \cos \theta-\left(c_{21} e^{i\varphi}+c_{2-1}e^{-i\varphi}\right) \sin \theta \neq 0\, ,
\eeq
implying that in the lowest multipole moment, at least quadrupolar anisotropies in the stress tensor of the fermion have to appear. The same results hold considering a right-handed fermion apart for a negative overall sign in the Boltzmann equations.

%%%%%%%%%%%%%%%%%%%%%%%%%%%%%%%%%%%%%%%%%%%%%%%%%%%

\section{General conditions for generating B-mode polarization}\label{Bmode}

As we have seen in the previous section, new interactions which are even or odd under party and even under time-reversal can generate V-modes, but are unable to generate B-mode polarization through the forward scattering term. This is due to the fact that in Eq. \eqref{Fijtensor} the term multiplying the $G_2$ coefficient vanishes being the amplitude even under time-reversal. Let us briefly explain this fact. After doing the expectation value of Eq. \eqref{forward}, the forward scattering contribution to the Boltzmann equations schematically reads as
\bea 
\frac{d\rho_{ij}(\mathbf K, \mathbf{k_c})}{dt} \propto i \, \int d\mathbf{q}\left(\delta_{is}\rho^{(\gamma)}_{s'j}(\mathbf{k_c})
-\delta_{js'}\rho^{(\gamma)}_{is}(\mathbf{k_c})\right) \delta_{rr'} \, n_f(\mathbf{q})
M^{r,r',s,s'}(\mathbf{q}'=\mathbf{q},\mathbf{p}= \mathbf{p}'=\mathbf{k}) \,,
\eea
where $M$ is the scattering amplitude of the process taken in the forward scattering limit. Now, we can express the Q-mode taking the difference between the $ij=11$ and $ij=22$ components of the polarization matrix. So, we have
\beq
  \frac{d}{dt}\, {Q}^{(S)}(\mathbf{K, k_c}) \propto i \, \int d\mathbf{q} \left[\left(\rho^{(\gamma)}_{s'1}(\mathbf{k_c})M^{r,r,1,s'}
-\rho^{(\gamma)}_{1s}(\mathbf{k_c}) M^{r,r,s,1}\right) - \left(\rho^{(\gamma)}_{s'2}(\mathbf{k_c})M^{r,r,2,s'}
- \rho^{(\gamma)}_{2s}(\mathbf{k_c}) M^{r,r,s,2} \right) \right] \, n_f(\mathbf{q}) 
 \,.
\eeq
Now, summing over the remaining $s$ and $s'$ indexes, the coupling with the U-modes is given by the following term:
\beq
  \frac{d}{dt}\, {Q}^{(S)}(\mathbf{K, k_c}) \propto i \, \int d\mathbf{q} \, {U}^{(S)}(\mathbf{K, k_c}) \, \left(M^{r,r,1,2}  - M^{r,r,2,1} \right) \, n_f(\mathbf{q}) 
 \,.
\eeq
From this last equation, we get that only scattering amplitudes that are antisymmetric in the final $s, s'$ photon polarization indexes can give a direct coupling between Q- and U-modes. This coupling converts E-modes directly into B-modes and vice versa. The only term in the amplitude of the process to have this property is the one proportional to the $G_2$ coefficient due to the Levi-Civita tensor contracting the photon polarization vectors in Eq. \eqref{Fijtensor}. All the other terms turn out to be symmetric in the $s$ and $s'$ indexes, thus not providing any direct coupling between the Q- and U-modes.

Now, in this section we will investigate the case in which the fermion-photon scattering amplitude is odd under time-reversal, leading for a non-negligible value of the $G_2$ term, thus providing a direct source term for B-mode polarization.

%%%%%%%%%%%%%%%%%%%%%%%%%%%%%%%%%%%%%%%%%%%%%%%%%%%%%%%%%%%%%%%%%%%%%%%%%%%%%%%%%%%%%%%%%%%%%%%%%%%%%%%%%%%%%%%%%%%%%%%%%

\subsection{Even-parity  and odd-time-reversal amplitude}
As we discussed in Sec. \ref{3}, after imposing the even-parity condition the $G_2$ coefficient is restricted to be [see Eqs.~(\ref{G02}) and~(\ref{Ceven})]
\beq
G_2=\bar{u}_{r'}(f_{11}\gamma^{5}+f_{12}\gamma^{5}\slashed{P})u_r\, .
\eeq
After imposing the odd-time-reversal condition, the only nonzero coefficients are
\beq
G_1=\bar{u}_{r'}(f_7\gamma^5)u_r\:\:\:\:\:\:\:\:\:\:\textrm{and}\:\:\:\:\:\:\:\:\:\:G_2=\bar{u}_{r'}(f_{12}\gamma^{5}\slashed{P})u_r\, .
\eeq
Moreover, we impose the odd charge conjugation condition, so that the amplitude is even under CPT. As a result we get $f_7=0$. Finally, the scattering amplitude is reduced to
 \beq
M_{fi}=-4f_{12}\, \bar{u}_{r'} \slashed{P} \gamma^5 u_r\, \frac{ \Delta^0}{m_f \sqrt{\mathbf{N}^2}}\,\left[|\mathbf q|^2\, \mathbf{p}\cdot(\bm{\epsilon}^{s}\times\bm{\epsilon}^{s'}) -  (\mathbf{q} \cdot \mathbf{p})  \,\left[\mathbf{q}\cdot(\bm{\epsilon}^{s}\times\bm{\epsilon}^{s'})\right] \right]\, .
\eeq
The only term which survives multiplies a factor of $\Delta^0$. Hence, the corresponding effect will be a loop quantum effect.
The time evolution of   the brightness Stokes parameters is given by
 \bea
  \frac{d}{d\eta}\, {Q}^{(S)}(\mathbf{K, k_c}) =- \frac{ a(\eta) f_{12}}{k^0_c  m_f}\,\int d\mathbf{q}\,n_{f}(\mathbf{K, q})\,\bar{u}_r \slashed{k}_c \gamma^5 u_r
      \, \frac{1}{\sin\psi}\left[\left(|\mathbf{ q}| \, \mathbf{\hat{k}_c}-  (\mathbf{\hat q} \cdot \mathbf{\hat k_c}) \, \mathbf{q}\right)\cdot (\bm{\epsilon_{1}}\times\bm{\epsilon_{2}})\right] U^{(S)}(\mathbf{K, k_c}) +\textrm{s.C.s.t.}\, ,\nonumber\\
\eea
and
\bea
  \frac{d}{d\eta}\, {U}^{(S)}(\mathbf{K, k_c})= \frac{a(\eta) f_{12}}{k^0_c m_f}\,\int d\mathbf{q}\,n_{f}(\mathbf{K, q})\,\bar{u}_r \slashed{k}_c \gamma^5 u_r \frac{1}{\sin\psi}\left[\left(|\mathbf{ q}| \, \mathbf{\hat{k}_c}-  (\mathbf{\hat q} \cdot \mathbf{\hat k_c}) \, \mathbf{q}\right)\cdot (\bm{\epsilon_{1}}\times\bm{\epsilon_{2}})\right]\,{Q}^{(S)}(\mathbf{K, k_c})
 +\textrm{s.C.s.t.}\, , \nonumber \\
\eea
and hence
\beq
  \frac{d}{d \eta}\, P^{\pm(S)}+i K\mu P^{\pm(S)}=\mp i \alpha' P^{\pm(S)}+\textrm{s.C.s.t.}\, ,\label{eqdeltap}
\eeq
where $\mu=\hat{k}_c \cdot \hat{K} = \cos \theta$, and $\alpha'$ is defined as
\beq \label{alpha}
\alpha'(\eta)= - \frac{a(\eta)  f_{12}}{k^0_c m_f}\,\int d\mathbf{q}\,n_{f}(\mathbf{K, q})\,\bar{u}_r \slashed{k}_c \gamma^5 u_r \frac{1}{\sin\psi}\left[\left(|\mathbf{ q}| \, \mathbf{\hat{k}_c}-  (\mathbf{\hat q} \cdot \mathbf{\hat k_c}) \, \mathbf{q}\right)\cdot (\bm{\epsilon_{1}}\times\bm{\epsilon_{2}})\right] \, ,
\eeq
with
\begin{equation}
\alpha(\eta)= -\int_\eta^{\eta_0} \alpha'(\eta')  \, d\eta'
\end{equation}
and
\bea
\sin\psi=[(\sin \theta \sin \theta' \sin(\varphi -\varphi'))^2+(\cos \varphi \cos \theta' \sin \theta- \cos\theta \cos \varphi' \sin \theta')^2+(\cos \theta' \sin \theta \sin \varphi - \cos \theta \sin \theta' \sin \varphi')^2]^{1/2} \, . \nonumber\\
\eea
As a result, Eq. \eqref{eqdeltap} can be rewritten as
\beq
   \frac{d}{d \eta}\left[{P}^{\pm (S)}\,e^{i K \mu\eta\, \pm\, i\alpha(\eta)\,-\tau(\eta)}\right]  = e^{iK \mu\eta\, \pm\, i\alpha(\eta)\,-\tau(\eta)}\left(
{1\over 2} \tau'\left[1-P_2(\mu)\right] \,\Pi\right)\, ,
\eeq
where again $\Pi= I^{2(S)} + P^{2(S)} - P^{0(S)}$.
Integrating the last equation gives the general solution
\begin{eqnarray}
{P}^{\pm (S)}
(\eta_0,{\mathbf K},\mu)&=&{3 \over 4}(1-\mu^2)\int_0^{\eta_0} d\eta\,
e^{iK(\eta - \eta_0) \mu \pm  i\alpha(\eta) -\tau(\eta)}\,\tau'(\eta)\,\Pi(\eta,{\bf K})\label{Boltzmann4} \, .
\end{eqnarray}
Then, using Eqs. \eqref{E}, \eqref{B}, and \eqref{P_to_EB}, we get the following expressions for the E- and B-modes
\beq
{E}^{(S)}(\eta_0,{\mathbf K},\mu)=
-\frac{3}{4}\int_{0}^{\eta_{0}}d\eta \,g(\eta)\,\Pi(\eta,{\mathbf K})\partial_{\mu}^{2} \left[(1-\mu^{2})^2e^{iK(\eta - \eta_0)\mu} \cos{\alpha(\eta)}\right]\, ,\label{newEmode}
\eeq
\beq
{B}^{(S)}(\eta_0,{\mathbf K},\mu)= - \frac{3}{4}\int_{0}^{\eta_{0}}d\eta \,g(\eta)\,\Pi(\eta,{\mathbf K})\partial_{\mu}^{2} \left[ (1-\mu^{2})^2e^{iK(\eta - \eta_0)\mu} \sin{\alpha(\eta)}\right]\, .\label{newBmode}
\eeq
where $g(\eta)=\tau'e^{-\tau}$ is the so-called visibility function.

Also in this case we need the fermion to be left- or right-handed, otherwise $\alpha = 0$ since  $\sum_r \bar{u}_r \gamma^\mu \gamma^5 u_r=0$. Anyway, in this case, the angular integral inside the definition of $\alpha$, Eq. \eqref{alpha}, is not equal to 0 if $n_{f}(\mathbf{K, q})$ is isotropic. Thus, we do not have to impose any particular condition to the fermionic stress tensor.

%%%%%%%%%%%%%%%%%%%%%%%%%%%%%%%%%%%%%%%%%%%%%%%%%%%%%%%%%%%%%%%%%%%%%%%%%%%%%%%%%%%%%%%%%%%%%%%%%%%%%%%%%%%%%%%%%%%%%%%%%%%%

\subsection{Odd-parity and odd-time-reversal amplitude}

The expressions of the coefficients $G_i$'s under odd-parity condition have been presented in Eq. \eqref{oddparity}.  Hence, $G_2$ is restricted to
\beq
G_2=\bar{u}_{r'}(f_9+f_{10}\slashed{P})u_r\, .
\eeq
Then applying the odd-time-reversal condition on $G_i$, we get
\beq
f_4=f_5=f_6=f_{16}=0\, .
\eeq
Therefore,
\bea
G_0=\bar{u}_{r'}(f_3\gamma^{5})u_r\, ,\:\:\:\:\:\:\:
G_1=0\, ,\:\:\:\:\:\:\:
G_2=\bar{u}_{r'}(f_9+f_{10}\slashed{P})u_r\, ,\:\:\:\:\:\:\:
G_3=\bar{u}_{r'}(f_{15}\gamma^{5})u_r\, ,\label{oddparity1}
\eea
which are all even under charge conjugation. Hence, the final form of amplitude will be even under CPT. Using these results, the final form of the amplitude is simplified to
\beq
M_{fi}= 4\frac{ \Delta^0}{m_f \sqrt{N^2}}\left(f_9+f_{10}\frac{P\cdot q}{m_f}\right) \left[|\mathbf q|^2\, \mathbf{p}\cdot(\bm{\epsilon}^{s}\times\bm{\epsilon}^{s'}) -  (\mathbf{q} \cdot \mathbf{p})  \,\left[\mathbf{q}\cdot(\bm{\epsilon}^{s}\times\bm{\epsilon}^{s'})\right] \right]\delta_{rr'}\, .
\eeq
The corresponding E-mode and B-mode polarizations are derived using the same method that we used to derive Eqs.~\eqref{newEmode} and \eqref{newBmode}. The only difference is that the parameter $\alpha'(\eta)$ changes into the following form:
 \beq
\alpha'(\eta)= \frac{a^2(\eta)}{ k^0_c m_f}\,\int d\mathbf{q}\,n_{f}(\mathbf{K, q})\left(f_9+2f_{10}\frac{q \cdot k_c}{ a(\eta) m_f}\right) \frac{1}{\sin\psi}\left[\left(|\mathbf{ q}| \, \mathbf{\hat{k}_c}-  (\mathbf{\hat q} \cdot \mathbf{\hat k_c}) \, \mathbf{q}\right)\cdot (\bm{\epsilon_{1}}\times\bm{\epsilon_{2}})\right]\, .
\eeq
As a result, in this case there is no restriction on the handedness of the fermion. In fact, $\alpha'$ can be different from zero if the fermion interacts both in the left- and right-handed states. Moreover, also in this case, we do not have to impose any particular condition in the fermion stress tensor since we do not need anisotropies for providing a value different from zero to the angular integral contained in the $\alpha'$ expression.
\begin{table}[htb]
\centering
\setlength{\tabcolsep}{15pt} % to act on column separation, default value: 6pt
\renewcommand{\arraystretch}{1.7} % to act on row separation, default value: 1	
\begin{tabular}{|c||c|c|}
	\hline
	Symmetries broken & V-mode formation & B-mode formation \\
	\hline
	All preserved & Anisotropies in $n_f({\mathbf K, q})$  & / \\
	\hline
	C and P &  \quantities{Anisotropies in $n_f({\mathbf K, q})$ \\ Only R- or L-handed fermion} & / \\
	\hline
	C and T & / & Only  R- or L-handed fermion \\
	\hline
	P and T & /& No conditions \\
	\hline
\end{tabular}
\caption{The conditions one needs to impose on the fermion to directly convert CMB E-modes into V- and B-modes through fermion-photon forward scattering in the different cases analyzed.} \label{table}
\end{table}
%%%%%%%%%%%%%%%%%%%%%%%%%%%%%%%%%%%%%%%%%%%%%%%%%%%%%%%%%%%%%%%%%%%%%%%%%%%%%%%%%%%%%%%%%%%%%%%%%%%%%%%%%%

\section{Majorana fermions} \label{7}

In the previous sections, we assumed the fermion to be a Dirac spinor. In this section, we will analyze what changes when the interacting fermion is a Majorana spinor, instead of a Dirac spinor. Analogous considerations have already been made in Ref. \cite{Xue:2014} for the case in which the fermion is a neutrino.

A Majorana fermion is a particle which coincides with its own antiparticle and hence it has no electric charge \cite{Mohapatra:2004,Giunt:2007,Akhmedov:2014kxa}.
The Majorana spinor is  defined as
\beq
\psi_M=\gamma^0 C \psi^{\ast}_M \, ,
\eeq
where $C$ is the charge conjugation operator. The properties of Majorana bilinear terms under parity, charge conjugation, and time-reversal transformations have been summarized in Refs. \cite{Mohapatra:2004,Giunt:2007,Akhmedov:2014kxa}. The Majorana condition implies $\psi_M=\psi^{c}_M$. As a result, a Majorana spinor transforms under charge conjugation as
\beq
C^{-1}\psi_M \,C=\psi_M\, .
\eeq
Thus, in general we can write
\beq
C^{-1}\left(\bar{\psi}_M A\, \psi_M \right)C=\bar{\psi}_M A\, \psi_M\, ,
\eeq
that for $A=\gamma^{\mu}$ becomes
\beq
\bar{\psi}_M \gamma^{\mu}\, \psi_M =0\, .
        \eeq
However, one can show that the transformations of the other Majorana bilinear terms under $P$, $T$, and $C$ are the same as Dirac bilinear terms. It was discussed in Ref. \cite{Latimer:2016kdg} that the Compton scattering amplitude for Majorana fermions is given by
\beq
M_{fi}=\bar{u}_{r'}(q')\epsilon^{s}_{\mu}\left[F^{\mu\nu}(q,q',p,p')+C\left(F^{\nu\mu}(-q',-q,p,p')\right)^TC^{-1}\right]\epsilon^{s'}_{\nu}u_{r}(q)\, ,
\eeq
$F^{\mu \nu}$ being as in Eq. \eqref{amplitude1}.
Now, if in general
\beq
C\left(F^{\nu\mu}(-q',-q,p,p')\right)^TC^{-1}=-F^{\mu\nu}(q,q',p,p')\, ,
\eeq
we find that $M_{fi}^{M}=0$ identically. However, if
\beq
C\left(F^{\nu\mu}(-q',-q,p,p')\right)^TC^{-1}=F^{\mu\nu}(q,q',p,p')\, ,
\eeq
then the scattering amplitude becomes
\beq
M^{M}_{fi}=2M^{D}_{fi}\, .
\eeq
Thus, when the Compton tensor $F^{\mu \nu}$ transforms like a pseudotensor under C, we get no fermion-photon forward scattering mixing. On the contrary, when $F^{\mu \nu}$ is invariant under C, we get the same coupling as discussed in the previous sections, but with an additional factor of 2 with respect to the Dirac fermion case.

%%%%%%%%%%%%%%%%%%%%%%%%%%%%%%%%%%%%%%%%%%%%%%%%%%%%%%%%%%%%%%%%%%%%%%%%%%%%%%%%%%%%%%%%%%%%%%%%%%%%%%%%%%%%%%%%%%%%%%%%
%%%%%%%%%%%%%%%%%%%%%%%%%%%%%%%%%%%%%%%%%%%%%%%%%%%%%%%%%%%%%%%%%%%%%%%%%%%%%%%%%%%%%%%%%%%%%%%%%%%%%%%%%%%%%%%%%%%%%%%%

\section{Conclusions} \label{8}

In the standard lore, circular and B-mode polarization of CMB photons cannot be generated via Compton scattering with electrons from linear scalar perturbations. In this work, we studied the conversion of CMB E-modes into V- and B-modes due to the forward scattering with a generic fermion in the presence of just linear scalar perturbations. We assumed interactions which may also go beyond the Standard Model of particle physics, keeping only gauge-invariance and the preservation of CPT symmetry. We derived various sets of Boltzmann equations describing the radiation transfer of CMB polarization. Our final results are qualitatively summarized in Table \ref{table}. We can have conversion in V-modes both preserving all the discrete symmetries and breaking the C and P symmetries. Instead, conversion into B-modes may arise only from the breaking of the T symmetry. Since our results are expressed in terms of free parameters, they offer a viable tool to put constraints on fundamental physics properties beyond the standard paradigms. An interesting extension of our work would be deriving the effects on CMB polarizations of the damping term in Eq. \eqref{Boltzmann equation} for the different interactions considered. We leave all these intriguing and interesting possibilities for future research.

\section*{Acknowledgements}

We would like to thank M. Kamionkowski and R. Mohammadi for useful comments and discussions. M.Z. would also like to thank INFN and the Physics and Astronomy Department ``G. Galilei" of Padova University for their warm hospitality during his visit in Padova. N.B., S.M., and G.O. acknowledge partial financial support by ASI Grant No. 2016-24-H.0.

\appendix

%%%%%%%%%%%%%%%%%%%%%%%%%%%%%%%%%%%%%%%%%%%%%%%%%%%%%%%%%%%%%%%%%%%%%%%%%%%%%%%%%%%%%%%%%%%%%%%%%%%%%%%%%%%%%%%%%%%%%%%%

\section{COMPARISON OF OUR AMPLITUDE WITH THE KIM AND DASS'S AMPLITUDE} \label{A}

In Ref. \cite{Kim:1975xf} Kim and Dass calculated the parity violating part of Compton amplitude using the procedure of Ref. \cite{Bardeen:1969aw}. They constructed $F^{\mu\nu}$ using the minimal pseudotensors violating parity. The general parity violating amplitude is defined as \cite{Kim:1975xf}
\bea
F^{\mu \nu}= \sum_i \mathcal{L}^i_{\mu \nu} A_i(x,y) \, ,
\eea
where $x=p\cdot q=p'\cdot q'$ and $y=p\cdot q'=p'\cdot q$. We change the kinematic variables defined in \cite{Kim:1975xf} to synchronize their notation with the notation of this paper. Moreover, we define a new variable $Q'$ as
\beq
Q'=q+q'~,
\eeq
and remove the factor $1/2$ adopted in \cite{Kim:1975xf} for kinematic variables. Hence, based on our notation, the
$\mathcal{L}^i_{\mu \nu}$ tensors defined in \cite{Kim:1975xf} are reconstructed as
\bea
\mathcal{L}^1_{\mu \nu}=Q'\cdot P\,\epsilon_{\mu \nu \alpha \beta} Q'^\alpha P^\beta + Q'_\mu N_\nu + N_\mu Q'_\nu\, ,
\eea
\bea
\mathcal{L}^2_{\mu \nu}=-\slashed{P}\left(Q'_\mu N_\nu + Q'_\nu N_\mu \right)+ Q'\cdot P \left(\gamma_\mu N_\nu + \gamma_\nu N_\mu \right)\, ,
\eea
\bea
\mathcal{L}^3_{\mu \nu}=\gamma^5 \slashed{P}\left( P^2 g_{\mu \nu} - P_\mu P_\nu + t_\mu t_\nu \right) \, ,
\eea
\bea
\mathcal{L}^4_{\mu \nu}=- P^2(Q'_\mu N_\nu + Q'_\nu N_\mu) + Q'\cdot P (N_\mu P_\nu + N_\nu P_\mu) \, ,
\eea
\bea
\mathcal{L}^5_{\mu \nu}=\gamma^5 \slashed{P} \left[ P^2 Q'_\mu Q'_\nu + (Q'\cdot P)^2 g_{\mu \nu} - Q'\cdot P (Q'_\mu P_\nu + Q'_\nu P_\mu)\right]\, .
\eea
We can express the $G_i$ coefficients defined in \eqref{amplitude1} in terms of the $A_i$ coefficients. The results are
\bea
G_0&=&\frac{1}{P^2}\left\{2\epsilon_{\mu \nu \rho \sigma} P^\mu Q^\nu t^\rho Q'^\sigma\left[ \left( A_1 - A_4 P^2\right)(P\cdot Q')^2 + P^2 \left(A_2 \slashed{P} +A_4 P^2 -A_1\right)\left(t^2 \left( (P\cdot Q)^2- P^2 Q^2\right)+Q'^2\right)\right] \right. \nonumber\\&&
 \left. + P^2\left[2A_2 t^2 (P\cdot Q')\left((P\cdot Q)^2 - P^2 Q^2\right) \gamma^5 \slashed{P}+\gamma^5 \slashed{P}\left[ (P\cdot Q')^2\left(A_5 t^2 (P\cdot Q)^2-A_5 Q'^2 -A_3\right) \right. \right.\right. \nonumber \\&&
 \left. \left. \left. +P^2 \left[t^2\left(A_5Q'^2 +A_3\right)(P\cdot Q)^2 - P^2 t^2 \left(Q^2 \left(A_5 Q'^2+A_3\right)- A_5(Q\cdot Q')^2\right)+ Q'^2(A_5Q'^2 + A_3)\right]  \right. \right.\right.\nonumber\\&&
 \left. \left. \left. - 2A_5 P^2 t^2 (P\cdot Q)(Q\cdot Q')(P\cdot Q')\right]\right] -2A_2 P^2 (P\cdot Q') \slashed{Q'} \epsilon_{\mu \nu \rho \sigma} P^\mu Q^\nu t^\rho Q'^\sigma\right\} \, ,
\eea\\
\bea
G_1&=& \frac{2}{P^2} \left\{ t^2 \left[(A_4 P^2 -A_1) \left[(P\cdot Q)^2\left((P\cdot Q')^2-2 P^2 Q'^2\right)-P^2\left(P^2 (Q\cdot Q')^2 + 2Q^2 \left((P\cdot Q')^2 -P^2 Q'^2\right)\right) \right. \right. \right.\nonumber\\
&& \left. \left. \left. +2P^2 (P\cdot Q)(Q\cdot Q')(P\cdot Q')\right]+2A_2P^2(P\cdot Q') \slashed{Q'} \left((P\cdot Q)^2-P^2 Q^2\right)+A_2 P^2 \slashed{Q} (P\cdot Q')\left(P^2 (Q\cdot Q') \right.\right.\right. \nonumber \\
&&\left.\left.\left. - (P\cdot Q)(P\cdot Q')\right)\right] + P^2 \slashed{P}\left[A_2 t^2\left((Q\cdot Q')\left[(P\cdot Q)(P\cdot Q')-P^2 (Q\cdot Q')\right]-2Q'^2 \left[(P.Q)^2 - P^2 Q^2\right]\right)\right.\right. \nonumber \\
&&\left.\left. -\gamma^5 P^2(A_5 Q'^2 +A_3) \epsilon_{\mu \nu \rho \sigma} P^\mu Q^\nu t^\rho Q'^\sigma\right]\right\} \, ,
\eea
\bea
G_2=0 \, ,
\eea
\bea
G_3&=& \frac{1}{P^2}\left\{ -2\epsilon_{\mu \nu \rho \sigma} P^\mu Q^\nu t^\rho Q'^\sigma \left[ P^2 (A_2 \slashed{P} +A_4 P^2 -A_1)\left( t^2\left(P^2 Q^2 - (P\cdot Q)^2\right)+Q'^2\right)+ (A_1-A_4 P^2)(P\cdot Q')^2\right]  \right. \nonumber\\
&& \left. + P^2 \left[2A_2 t^2 \gamma^5 \slashed{P} (P\cdot Q') \left((P\cdot Q)^2- P^2 Q^2\right)+ \gamma^5 \slashed{P} \left[(P\cdot Q')^2\left(A_5 t^2 (P\cdot Q)^2 + A_5 Q'^2 +A_3\right) \right. \right. \right. \nonumber \\
&& \left. \left. \left. - P^2 \left[-t^2(A_5 Q'^2 +A_3)(P\cdot Q)^2 +P^2 t^2\left(Q^2(A_5Q'^2 +A_3)-A_5 (Q\cdot Q')^2\right)+ Q'^2 (A_5 Q'^2 +A_3)\right] \right. \right. \right.\nonumber\\
&& \left. \left. \left. -2A_5 P^2 t^2 (P\cdot Q)(Q\cdot Q')(P\cdot Q')\right]\right]+2A_2 P^2 (P\cdot Q') \slashed{Q'} \epsilon_{\mu \nu \rho \sigma} P^\mu Q^\nu t^\rho Q'^\sigma \right\} \, .
\eea

As one can see, $G_2=0$ is consistent with what had been found in Eq. \eqref{Gcoefficientsodd}.

\section{INTERACTION OF PHOTONS WITH NEUTRINO MAGNETIC MOMENT} \label{loop}

To be able to consistently define the basis vector $\hat e^{(1)\lambda}$ in the forward scattering limit, Eq. \eqref{e1}, we introduced a new variable $\Delta^{\lambda}$, which takes the place of $t^{\lambda}$ in the general definition \eqref{N} and claimed that the subsequent new terms in the Compton tensor may arise from loop corrections in the Feynman diagrams. Here, we consider an explicit example and compare the Compton tensor of this example with the general forward scattering Compton tensor derived at the end of Sec. \ref{4}.

If the neutrino has a magnetic moment, its interaction with photon is characterized by the following effective Hamiltonian \cite{Karl:2004bt,Mohanty:1997mr,Royer:1968rg} (for a recent review, see Ref. \cite{Giunti:2014ixa}):
\beq
\mathcal{H}\sim\mu_e\mu_{\nu}\bar{u}_{\nu}(q')\sigma_{\alpha\beta}u_{\nu}(q)\mathcal{F}^{\alpha\beta}\, ,\label{Hintn}
\eeq
where $\mathcal{F}^{\alpha\beta}$ is the field strength of the photon, $\mu_e$ is the magnetic moment of the electron and $\mu_{\nu}$ is the magnetic moment of the neutrino. From Eq. \eqref{Hintn}, we can derive the forward scattering amplitude in the following form
\beq
M_{fi}=F^{\lambda\tau}\epsilon^{s'\ast}_{\lambda}\epsilon^s_{\tau}\, ,
\eeq
where
\beq
F^{\lambda\tau}\sim (\mu_e\mu_{\nu})^2\epsilon^{\lambda\tau\alpha\beta}p_{\alpha}q_{\beta}\, ,
\eeq
$p$ being the photon momentum and $q$ the neutrino momentum.

It is immediate to verify that this term is equivalent in form to the second term of Eq. \eqref{G2_term}, which contains the quantity $\Delta^{\lambda}$. This simple example shows that effectively the new definition of $\hat e^{(1)\lambda}$ is sensitive to loop quantum effects and provides a more general expression for the Compton tensor that, as discussed in Sec. \ref{Bmode}, may cause
B-mode generation in the CMB.

\section{FORWARD SCATTERING MIXING INDUCED BY P AND C BREAKING INTERACTIONS} \label{app:C} 

Here we present the full expressions of the forward scattering polarization mixing terms in the Boltzmann equations obtained in the case in which parity symmetry is broken (Sec. \ref{odd_parity}). In the Dirac representation, the helicity spinors are given by \cite{Itzykson:1980rh}
\beq
 u_{R} (\mathbf q)=\sqrt{\frac{q^0+m_f}{2m_f}}\left(
                   \begin{array}{cc}
                    \cos(\frac{\theta'}{2})  \\\\
                    \sin(\frac{\theta'}{2}) e^{i\varphi'} \\\\
                    \frac{\mathbf{|q|}}{q^0+m_f} \cos (\frac{\theta'}{2})\\\\
                     \frac{\mathbf{|q|}}{q^0+m_f} \sin (\frac{\theta'}{2}) e^{i\varphi'}\\\\
                   \end{array}
                 \right)\, ,\:\:\:\:\:\:\:\:\:\:\:\:\:\:\:
 u_{L} (\mathbf q)=\sqrt{\frac{q^0+m_f}{2m_f}}\left(
                   \begin{array}{cc}
                   -\sin(\frac{\theta'}{2})   \\\\
                    \cos(\frac{\theta'}{2}) e^{i\varphi'} \\\\
                    \frac{\mathbf{|q|}}{q^0+m_f} \sin (\frac{\theta'}{2})\\\\
                   -\frac{\mathbf{|q|}}{q^0+m_f} \cos (\frac{\theta'}{2}) e^{i\varphi'}\\\\
                   \end{array}
                 \right)\, .
\eeq
Hence, the bilinear term $\bar{u} \gamma^\mu \gamma^5 u$ reads
\beq
\bar{u} \gamma^\mu \gamma^5 u= - \frac{1}{m} \left( |\mathbf{q}|,  q^0 \sin\theta' \, \cos \varphi', q^0 \sin\theta' \, \sin \varphi', q^0  \cos \theta'\right)\, .\label{bilinear term}
\eeq
Now, using Eqs. \eqref{nexpand}, \eqref{bilinear term}, and integrating over the fermion spherical angles $\varphi'$ and $\theta'$, Eqs. \eqref{E-SPQ}-\eqref{E-SPV} become
 \beq\begin{split}
   \frac{d}{d\eta}\, {Q}^{(S)}(\mathbf{K, k_c})=& i \frac{\sqrt{\pi}}{420}\frac{a(\eta)\,f_{\textrm{p}}}{ 16\pi^3  m^2_f}\,\int d|\mathbf{q}| \frac{ |\mathbf{q}|^4}{q^0}n_f(\mathbf{K}, |\mathbf{q}|)  \,\left\{ -28\sqrt{30}|\mathbf{q}| \left[\left(c_{22} e^{2i\varphi}-c_{2-2} e^{-2i\varphi}\right) \cos \theta+\left(c_{21} e^{i\varphi}+c_{2-1}e^{-i\varphi}\right)  \right.\right. \\
& \left.\left. \times \sin \theta\right]+10\sqrt{21} q^0 \sin (2\theta)\left(c_{31} e^{i\varphi}+c_{3-1}e^{-i\varphi}\right)+4\sqrt{210} q^0 \cos (2\theta)\left(c_{32} e^{2i\varphi}-c_{3-2}e^{-2i\varphi}\right) \right. \\
& \left.-6\sqrt{35} q^0 \sin (2\theta)\left(c_{33} e^{3i\varphi}+c_{3-3}e^{-3i\varphi}\right) \right\}\,{V}^{(S)}(\mathbf{K,k_c}) +\,\,\textrm{s.C.s.t.}\, ,
\end{split}
\eeq
\beq \begin{split}
 \frac{d}{d\eta}\, {U}^{(S)}(\mathbf{K, k_c})=&-
 \frac{\sqrt{\pi}}{420}\frac{a(\eta)\,f_{\textrm{p}}}{ 32 \pi^3  m^2_f}\,\int d|\mathbf{q}| \frac{ |\mathbf{q}|^4}{q^0} n_f(\mathbf{K}, |\mathbf{q}|)  \left\{-168\sqrt{5} |\mathbf{q}| \sin^2\theta\, c_{20}-14\sqrt{30} |\mathbf{q}| \left[2\left(c_{21}e^{i\varphi}-c_{2-1} e^{-i\varphi}\right) \sin 2\theta \right. \right.\\
&  \left.\left.+\left(c_{22} e^{2i\varphi}+c_{2-2} e^{-2i\varphi}\right)( \cos 2\theta+3) \right]+120\sqrt{7} q^0 \cos \theta \sin^2 \theta\, c_{30}+10\sqrt{21} q^0 \left(c_{31}e^{i\varphi}-c_{3-1} e^{-i\varphi}\right)\right.\\
&\left. \times \sin \theta (3\cos 2\theta +1) + \sqrt{210} q^0 \left(c_{32} e^{2i\varphi}+c_{3-2} e^{-2i\varphi}\right)\left(5 \cos \theta +3\cos 3\theta\right) \right.\\
& \left. - 6\sqrt{35}q^0 \left(c_{33} e^{3i\varphi}-c_{3 -3} e^{-3i\varphi}\right) \sin \theta (\cos 2\theta +3)\right\} \,{V}^{(S)}(\mathbf{K, k_c})+\textrm{s.C.s.t.}\, ,
\end{split}
\eeq
\beq \begin{split}
\frac{d}{d\eta}\, {V}^{(S)}(\mathbf{K, k_c})&= \frac{\sqrt{\pi}}{420}\frac{a(\eta)\,f_{\textrm{p}}}{ 32 \pi^3  m^2_f}\,\int d|\mathbf{q}| \frac{ |\mathbf{q}|^4}{q^0} n_f(\mathbf{K}, |\mathbf{q}|)
\left[-2i\left\{ -28\sqrt{30} |\mathbf{q}| \left[\left(c_{22} e^{2i\varphi}-c_{2-2} e^{-2i\varphi}\right) \cos \theta+\left(c_{21} e^{i\varphi}+c_{2-1}e^{-i\varphi}\right) \right. \right.\right. \\
& \left. \left. \left.  \times \, \sin \theta\right] +10\sqrt{21} q^0 \sin (2\theta)\left(c_{31} e^{i\varphi}+c_{3-1}e^{-i\varphi}\right)+4\sqrt{210} q^0 \cos (2\theta)\left(c_{32} e^{2i\varphi}-c_{3-2}e^{-2i\varphi}\right)-6\sqrt{35} q^0 \sin (2\theta)\right.\right. \\
&\left.\left.\times\left(c_{33} e^{3i\varphi}+c_{3-3}e^{-3i\varphi}\right) \right\}  {Q}^{(S)}(\mathbf{K, k_c})  - \left\{168\sqrt{5} |\mathbf{q}| \sin^2\theta\, c_{20}+14\sqrt{30} |\mathbf{q}| \left[2\left(c_{21}e^{i\varphi}-c_{2-1} e^{-i\varphi}\right) \sin 2\theta \right.\right.\right.\\
&\left.\left.\left.+\left(c_{22} e^{2i\varphi}+c_{2-2} e^{-2i\varphi}\right)( \cos 2\theta+3) \right]-120\sqrt{7} q^0 \cos \theta \sin^2 \theta\, c_{30}-10\sqrt{21} q^0 \left(c_{31}e^{i\varphi}-c_{3-1} e^{-i\varphi}\right) \sin \theta \right.\right.\\
& \left.\left. \times (3\cos 2\theta +1) -\sqrt{210} q^0 \left(c_{32} e^{2i\varphi}+c_{3-2} e^{-2i\varphi}\right)\left(5 \cos \theta +3\cos 3\theta\right)
 +6\sqrt{35}q^0 \left(c_{33} e^{3i\varphi}-c_{3 -3} e^{-3i\varphi}\right) \sin \theta \right.\right.\\
&\left. \left. \times (\cos 2\theta +3)\right\}{U}^{(S)}(\mathbf{K, k_c})\right] \, +\,\,\textrm{s.C.s.t.}\, .~~
\end{split}
\eeq

%===============================================================%
%************************* BIBLIOGRAPHY ************************%
%===============================================================%
\begingroup %------------------------------ BIBLIOGRAPHY
\makeatletter
\let\ps@plain\ps@empty
\makeatother
\bibliography{bibl}
\endgroup
\end{document}